\documentclass[useAMS,usenatbib]{mnras}
\usepackage{times}
\usepackage{natbib}
\usepackage{amssymb,amsmath}
\usepackage{graphicx,color}
\usepackage{hyperref}
\usepackage[usenames,dvipsnames]{xcolor}
\usepackage[OT2,T1]{fontenc}
\usepackage[draft]{todonotes}
\usepackage{multirow}
\usepackage{mathtools}
\usepackage[normalem]{ulem}
\usepackage{lineno}
\usepackage{soul}
\newcommand{\om}{\Omega_\mr m}
\newcommand{\mr}{\mathrm}
\newcommand{\pd}{P_{\delta}}

\newcommand{\hdelta}{\ensuremath{\widehat{\kappa}}}
\newcommand{\hxi}{\ensuremath{\widehat{\xi}}}
\newcommand{\mean}[1]{\ensuremath{\left\langle #1 \right\rangle}}
\renewcommand{\vec}[1]{\mathbf{#1}}
\newcommand\orout{\bgroup\markoverwith
{\textcolor{orange}{\rule[.5ex]{2pt}{1pt}}}\ULon}
\title[Non-Gaussian likelihood]{Non-Gaussianity in the Weak Lensing Correlation Function Likelihood - Implications for Cosmological Parameter Biases}

\author[C.-H. Lin et al.]
{\parbox{\textwidth}{Chien-Hao Lin$^1$\thanks{E-mail: \texttt{chienhal@andrew.cmu.edu}},
Joachim Harnois-D\'eraps$^2$,
Tim Eifler$^{3,4}$,
Taylor Pospisil$^5$,
Rachel Mandelbaum$^1$,
Ann B. Lee$^5$, and
Sukhdeep Singh$^6$
\\(The LSST Dark Energy Science Collaboration)}
\vspace{0.4cm}\
\\$^1$McWilliams Center for Cosmology, Department of Physics, 
Carnegie Mellon University, Pittsburgh, PA 15213, USA,
\\$^2$Scottish Universities Physics Alliance, Institute for Astronomy, University of Edinburgh, Royal Observatory, Blackford Hill, Edinburgh, EH9 3HJ, U.K.,
\\$^3$Jet Propulsion Laboratory, California Institute of Technology, Pasadena, CA 91109 USA,\ 
\\$^4$Steward Observatory/Department of Astronomy, University of Arizona, 933 North Cherry Avenue, Tucson, AZ 85721, USA\ 
\\$^5$Department of Statistics \& Data Science, Carnegie Mellon University, Pittsburgh, PA 15213, USA,
\\$^{6}$ Berkeley Center for Cosmological Physics, Department of Physics, \& Lawrence Berkeley National Laboratory,\\ University of California, Berkeley, CA 94720, USA
}

\begin{document}
\date{\today}
\maketitle

\begin{abstract}
We study the significance of non-Gaussianity in the likelihood of weak lensing shear two-point correlation functions, detecting significantly non-zero skewness and kurtosis in one-dimensional marginal distributions of shear two-point correlation functions in simulated weak lensing data. We examine the implications in the context of future surveys, in particular LSST, with derivations of how the non-Gaussianity scales with survey area. 
We show that there is no significant bias in one-dimensional posteriors of $\Omega_{\rm m}$ and $\sigma_{\rm 8}$ due to the non-Gaussian likelihood distributions of shear correlations functions using the mock data ($100$ deg$^{2}$).
We also present a systematic approach to constructing approximate multivariate likelihoods with one-dimensional parametric functions by assuming independence or more flexible non-parametric multivariate methods after decorrelating the data points using principal component analysis (PCA). 
While the use of PCA does not modify the non-Gaussianity of the multivariate likelihood, we find empirically that the one-dimensional marginal sampling distributions of the PCA components exhibit less skewness and kurtosis than the original shear correlation functions.
Modeling the likelihood with marginal parametric functions based on the assumption of independence between PCA components thus gives a lower limit for the biases.
We further demonstrate that the difference in cosmological parameter constraints between the multivariate Gaussian likelihood model and more complex non-Gaussian likelihood models would be even smaller for an LSST-like survey. In addition, the PCA approach automatically serves as a data compression method, enabling the retention of the majority of the cosmological information while reducing the dimensionality of the data vector by a factor of $\sim$5.
\end{abstract}


\section{Introduction}

\label{sec:intro}
Weak gravitational lensing is the deflection of light by the gravitational field of large-scale structure, which leads to minute distortions of the observed galaxy images compared to their original shapes in the galaxy source plane. Measuring the correlation functions of the galaxy shapes is therefore a way to measure the growth of structure and the geometry of the Universe \citep[e.g.,][]{2001PhR...340..291B, 2008ARNPS..58...99H, 2015RPPh...78h6901K, doi:10.1146/annurev-astro-081817-051928} and hence a promising avenue to constrain cosmology \citep[][]{ heh14,2016ApJ...824...77J, 2018arXiv181206076H, PhysRevD.98.043528,2019PASJ...71...43H}.  

With the next generation of weak lensing surveys, such as the Large Synoptic Survey Telescope (LSST\footnote{http://www.lsst.org/lsst}; \citealt{2019ApJ...873..111I}), the Wide-Field Infrared Survey Telescope (WFIRST\footnote{https://wfirst.gsfc.nasa.gov}) and \textit{Euclid}\footnote{http://sci.esa.int/euclid/}, we expect data sets that are both wider and deeper compared to current surveys (e.g.\ KiDS\footnote{http://kids.strw.leidenuniv.nl}, DES\footnote{https://www.darkenergysurvey.org}, HSC\footnote{http://hsc.mtk.nao.ac.jp/ssp/}) in the near future. The high-quality data from upcoming lensing surveys is expected to reduce the statistical uncertainty in weak lensing measurements compared to current surveys by an order of magnitude. In order to fully exploit the cosmological constraining power of weak lensing surveys, much effort has been committed to understand observational systematic effects, such as uncertainties in the measurement of galaxy shapes \citep[e.g.,][]{doi:10.1093/mnras/sts371, doi:10.1093/mnras/stv781} and photometric redshifts. On the astrophysical side, major progress has been made in modeling systematics that affect the interpretation of the weak lensing signal, such as the intrinsic alignment of galaxies \citep[e.g.,][]{2011MNRAS.410..844M,2015SSRv..193....1J,2015PhR...558....1T, 2016MNRAS.456..207K}, the nonlinear evolution of the dark matter density field \citep[e.g.,][]{tsn12,2014ApJ...780..111H}, and baryonic effects that modify the latter \citep[e.g.,][]{2015MNRAS.454.2451E,2015MNRAS.454.1958M,2018MNRAS.480.3962C}. 

However, inaccuracies in the last step of the analysis, the inference of cosmological parameters from measurements of observables such as $\xi_{\pm}$ and some model for the likelihood function, are less well-explored in the weak lensing community. Uncertainties related to parameter space sampling, discrepancy metrics, and the likelihood of the summary statistics itself are important aspects of the cosmological interpretation that can lead to potential biases in the analysis. 

While likelihood-free approaches such as Approximate Bayesian Computation \citep[e.g.,][]{2015JCAP...08..043A,2017A&A...599A..79P} are beginning to emerge as a tool for cosmological inference, most analyses still assume a likelihood function to transition from observations to cosmological parameters. Among the possible choices, the multivariate Gaussian likelihood function is the simplest and most commonly used. 

Even though the Cosmic Microwave Background (CMB) temperature field is close to a Gaussian and the non-Gaussian features in CMB power spectra are small, the CMB data analyses \citep{2013ApJS..208...19H, 2014A&A...571A..15P,2016A&A...594A..13P} have made some progress beyond this simple Gaussian assumption.

They use second-order statistics to capture the cosmological information in the underlying temperature and polarization field, which are (very close to) Gaussian. For idealized CMB observations (full-sky, isotropic beam, spatially uniform noise) the empirical power spectra of the underlying Gaussian temperature and polarization fields has a 
Wishart distribution given the model power spectra. At sufficiently high $\ell$, the likelihood function of power spectra approaches a multivariate Gaussian following the Central Limit Theorem. When going beyond the idealized case, the inclusion of potentially non-Gaussian foreground distributions such as galactic dust emission, Cosmic Infrared Background, and radio point sources breaks the initial assumption that the measured field is Gaussian and consequently breaks the conclusion that the likelihood of the power spectra beyond a certain $\ell$ is well-approximated by a multivariate Gaussian. As further detailed in \cite{2016A&A...594A..13P}, the Planck analysis masks foreground contaminants and assumes that the non-Gaussian features are subdominant outside of the masked regions.

The situation is different for weak lensing. Due to non-linear structure evolution at late times, the shear field itself is non-Gaussian, which invalidates the premise of the CMB argument. Nevertheless, weak lensing analyses have assumed a multivariate Gaussian likelihood function to be the underlying distribution of shear two-point statistics \citep{2008A&A...479....9F, doi:10.1111/j.1365-2966.2010.17430.x, heh14, doi:10.1093/mnras/stw2805, PhysRevD.98.043528}. In these analyses, the non-Gaussianity of the shear field enters only via non-linear matter density power spectra \citep{tsn12,2014ApJ...780..111H} that are used to model the observed two-point statistics and via so-called non-Gaussian covariances, which indicate that a non-vanishing four-point function is included in the covariance computation. 

Non-Gaussian shear covariances due to non-linear clustering have been studied in \cite{2009MNRAS.395.2065T,2009ApJ...701..945S,2011ApJ...734...76S, 2015MNRAS.450.2857H}. These studies use the multivariate Gaussian shear likelihood with the contribution of covariance from non-Gaussian fields included.  The impact on cosmological constraints depends on the scales considered and on depth and area of the survey. 
Computing non-Gaussian covariances is essential in cosmic shear analyses, and most recent cosmic shear measurements have opted for different strategies, such as analytical computation of Gaussian and non-Gaussian covariances \citep[e.g.,][]{2016ApJ...824...77J,2017MNRAS.465.1454H,2017arXiv170609359K}, or for covariance estimated through numerical simulations \citep[e.g.][]{2013MNRAS.432.2433H}. For future surveys, several studies indicate the high computational costs of a brute-force numerical simulations approach \citep{2013PhRvD..88f3537D, 2014MNRAS.442.2728T}, which led to the development of new covariance estimators \citep{2017MNRAS.466L..83J,2018MNRAS.473.4150F}.  For cosmic shear, \cite{2018JCAP...10..053B} have shown that the Gaussian covariance plus Super Sample Covariance terms are sufficient, both of which can be easily implemented analytically. 

When going beyond the Gaussian likelihood function for the convergence power spectrum, previous studies have considered a lognormal distribution and the copula method for describing the non-Gaussian distribution \citep{2002ApJ...571..638T, 2011A&A...536A..85H,2011PhRvD..83b3501S, 2010PhRvL.105y1301S}. In configuration space, \citet{2009A&A...504..689H} revisited the assumption of Gaussianity for the two-point correlation function and tested the non-Gaussianity with independent component analysis (ICA). The authors measure the distribution of the correlation function in 9600 realizations of ray-tracing simulations resembling the Chandra Deep Field South lensing analysis and perform three full likelihood analyses: using ICA, a standard multivariate Gaussian with a ray-tracing covariance, and with a Gaussian covariance. This paper triggered several attempts to build analytical expressions for the likelihood function of the shear two-point correlation function that improve over the standard Multivariate Gaussian approximation. For example, \cite{2009A&A...504..705S} have shown that two-point correlation functions of Gaussian fields cannot take arbitrary values since this would violate the constraint of non-negativity of power spectrum. As a consequence the sampling distribution of the correlation functions cannot be an exact multivariate Gaussian. \cite{Keitel2011} employ Fourier mode expansion and characteristic functions of a Gaussian random field to derive an analytical expression for the likelihood function of its uni- and bi-variate correlation functions. In \cite{2013A&A...556A..70W, 2015A&A...582A.107W}, the authors transform the correlation functions such that a quasi-Gaussian approximation of the likelihood function is justified and tested its performance with simulations. A recent paper by \cite{2018MNRAS.473.2355S} explored the high-order correlations between the data points of the CFHTLenS cosmic shear correlation functions in search for non-Gaussianity. In \cite{2018MNRAS.477.4879S}, the authors measure the skewed distributions of weak lensing shear correlation functions in simulations and follow the CMB literature in developing an analytical expression for the correlation function likelihood. 

It is well-established in the literature that the sampling distribution of shear two-point correlation functions is not strictly Gaussian. Despite this, the Gaussian likelihood model is still the standard in weak lensing likelihood analysis for current surveys. If the likelihood is not Gaussian, then analysing the data with a Gaussian assumption could bias the cosmological parameter constraints. It is therefore important to quantify potential cosmological parameter biases that may arise due to un-modeled aspects of the likelihood function, so as to determine whether an alternative way of modeling the likelihood function is needed. This effect was discussed in the literature (e.g. \citealt{2009A&A...504..689H, 2018MNRAS.477.4879S, 2019PhRvD.100b3519T}), but agreement on whether the Gaussian approximation cause a significant bias for cosmological weak lensing analyses has not yet been reached. It is difficult to fully demonstrate the impact of the non-Gaussian likelihood on cosmological parameter constraints. Part of the difficulty comes from the dimensionality of the problem, with reconstructing in detail the full high-dimensional non-Gaussian likelihood currently being an unsolved problem. 
In this work, we do not address the full problem, but make some advance upon previous work by modelling the non-Gaussian likelihood with a large ensemble of weak lensing simulations, and further estimate the biases on cosmological parameters.

This paper is structured as follows: In Sect.~\ref{sec:sim}, we describe the details of our simulated weak lensing data. In Sect.~\ref{sec:method}, after showing how the data vectors are modeled theoretically, we describe the likelihood analysis and likelihood models. Section~\ref{sec:model_assessment} expands on the systematic approach of assessing the performance of likelihood models and data compression. In Sect.~\ref{sec:result}, we show the results for non-Gaussianity of weak lensing observables and the impact on cosmological parameter estimates.  Section~\ref{sec:conclusion} contains our discussion and conclusions.

\section{Simulations}
\label{sec:sim}
The simulated (mock) weak lensing data that are used in this paper are based on the Scinet LIght Cone Simulations\footnote{https://slics.roe.ac.uk/} \citep[][SLICS hereafter]{2015MNRAS.450.2857H, 2018MNRAS.481.1337H}, which are specifically tailored for statistical studies of weak lensing analyses. They consist of a series of lines-of-sight (LOS) of 100 deg$^2$ each, constructed by ray-tracing in their own independent realization. In the simulations, no masks are applied, and hence the patches are 10 by 10 deg$^2$. The underlying $N$-body simulations evolved $1536^3$ dark matter particles in a box length of 505 $h^{-1}$Mpc, and produced 18 mass planes between redshift 0.0 and 3.0, which are then converted into shear maps using the Born approximation. We used 932 such independent realizations in which the initial random seeds changed prior to the $N$-body run, with the assumed cosmology fixed to that of WMAP9+SN+BAO flat $\Lambda$CDM cosmology \citep{2013ApJS..208...19H}: $\Omega_{\rm m} = 0.2905$, $\Omega_\Lambda = 0.7095$, $\Omega_{\rm b} = 0.0473$, $h = 0.6898$, $\sigma_8 = 0.826$ and $n_s = 0.969$.  

The mock galaxy catalogues are then created in a way that is meant to reproduce the redshift distributions of weak lensing source galaxies in LSST \citep{2013MNRAS.434.2121C}: 
\begin{eqnarray}
n(z) \propto z^\alpha \exp\left[{-\left(\frac{z}{z_0}\right)^\beta}\right],
\label{eq:nz}
\end{eqnarray}
with $\{\alpha, \beta, z_0\} = \{1.21, 1.0, 0.5\}$, assuming a source number density of 26 gal/arcmin$^2$. The mocks are split in 10 tomographic redshift bins $n_i(z)$, each containing the same number of galaxies. These distributions are further smoothed by a Gaussian kernel of width $\sigma = (1+z)\sigma_z$ and $\sigma_z = 0.02$. For the detailed redshift distributions of the 10 LSST-like source bins, see Fig.~A1 in \cite{2018MNRAS.481.1337H}.

Besides the cosmological shear $\gamma$ (see Sect. \ref{sec:data-vector}), the observed ellipticity $\epsilon^{\rm obs}$ includes the intrinsic shapes of galaxies $\epsilon_{\rm int}$ through the shear addition formula:
\begin{eqnarray}
\epsilon^{\rm obs} = \frac{\gamma  + \epsilon_{\rm int}}{1 + \gamma \epsilon_{\rm int}^{*}}.
\label{eq:eps_noise}
\end{eqnarray}
In the above expression, shear and ellipticities are written as complex variables, and the $\epsilon^{\rm obs}_{1/2}$ components are recovered from the real and imaginary parts respectively. 
The two components of the intrinsic galaxy shapes are each drawn from a Gaussian distribution with zero mean, a standard deviation of 0.29 inspired by the KiDS-450 \citep{doi:10.1093/mnras/stw2805} data, and the constraint that $|\epsilon_{\rm int}|^2 \le 1$.

In our analysis, we measure the non-Gaussian shapes of the likelihood function of shear correlation functions with the SLICS simulation, and then use them to estimate the parameter biases for large weak lensing surveys such as LSST. Due to the differences in survey areas, different sources of uncertainty dominate: the 100 deg$^2$ SLICS simulations are shape-noise dominated while LSST data will be cosmic-variance dominated. 
Therefore, in this work we sometimes switch off the intrinsic shape noise in order to separately understand the contributions of cosmic variance and shape noise to the shape of the likelihood function. For all results that are presented, we refer to the results as `without shape noise' or `with shape noise'.

\section{Method}
\label{sec:method}
\subsection{Cosmic shear correlation function data vector}
\label{sec:data-vector}
The weak lensing effect is mathematically approximated as a linear transformation that maps the unlensed location to the lensed location. The transformation matrix $\mathcal{A}$, which connects the shape of a source with the observed images, can be written as 
\begin{equation}
\mathcal{A} = \left(\begin{array}{cc} 1-\kappa-\gamma_1 & -\gamma_2\\ -\gamma_2 & 1-\kappa+\gamma_1 \end{array}\right),
\end{equation}
where $\kappa$ is the convergence and $\gamma_1, \gamma_2$ are the two components of the spin-two shear $\gamma = \gamma_1 + \mathrm{i}\gamma_2$ in Cartesian coordinates. Conventionally the coordinate system is rotated so that the separation vector is parallel to the $x$-axis. The shear components are decomposed into the tangential direction ($\gamma_t$) and the cross direction ($\gamma_\times$) in the rotated coordinates:
\begin{equation}
\gamma_t = - \mathbf{Re}(\gamma e^{-2\mathrm{i}\phi}), \gamma_\times = - \mathbf{Im}(\gamma e^{-2\mathrm{i}\phi}),
\end{equation}
where $\phi$ is the polar angle of the separation vector of the two galaxies. 
From the shear components we can write the two shear correlation functions as a function of angular separation
\begin{equation}
\xi_\pm^{ij}(\theta) = \left\langle \gamma_t^{i}\gamma_t^{j} \right\rangle(\theta) \pm \left\langle \gamma_\times^{i}\gamma_\times^{j} \right\rangle(\theta) .
\end{equation}
Here $i$ and $j$ are indices of tomographic redshift bins, and the angle brackets refer to ensemble average. The correlation functions are computed from the SLICS simulations with the package \textsc{TreeCorr}\footnote{https://github.com/rmjarvis/TreeCorr} \citep{2004MNRAS.352..338J}.

The angular bins for measuring shear correlation functions initially divide the scales from 0.32 to 400 arcmin into logarithmically spaced bins of width 
$\Delta({\rm ln}\theta)=0.23$; however, we further apply angular selections to minimize the impact of known limitations in the simulations \citep{2015MNRAS.450.2857H}.  We require $\theta> 0.8$ arcmin for $\xi_+$, to avoid resolution effects in the simulations at small scales. Since $\xi_-$ is more sensitive to small-scale uncertainties, we apply a more aggressive constraint: $\theta > 6.5$ arcmin. We also require $\theta<160$ arcmin for $\xi_+$ to avoid scales with significant power loss due to the box size. On the other hand, $\xi_-$ is less affected by the box size effect within 400 arcmin; therefore we do not introduce extra constraints on the large scale for $\xi_-$.  

These scales are similar to those used in the recent KiDS-450 cosmic shear analysis by Hildebrandt et al. (2017), and presume that the analysis has separately accounted for uncertainties at small scales due to baryon feedback, modeling of the non-linear power spectrum and of the intrinsic alignments of galaxies\footnote{Note that these effects could contribute to the level  of non-Gaussianity in the data, but that is beyond the scope of this paper.}. 
Modeling these uncertainties at the precision required by LSST will be challenging. In the recent DES analysis \citep{tmz17} conservative scale cuts were implemented in order to avoid biases due to imperfect modeling of said astrophysical effects.
For this paper, we assume an optimistic scenario in which the LSST weak lensing measurement pipeline includes these small angular scales. 

The complete data vector is the concatenation of all $\xi_+$ and $\xi_-$ values across all tomographic bins and $\theta$ bins. Each simulated realization has 55 correlation functions, each with 42 angular bins (24 for $\xi_+$ and 18 for $\xi_-$).

The estimator of the data covariance from the many simulation ensemble is defined as
\begin{equation}
\label{eq:covariance}
\hat{\mathbf{C}}^{ij} = \frac{1}{\nu} \sum_{k}^{N_s}{(\xi_k^{ij} - \bar{\xi}^{ij}) (\xi_k ^{ij}- \bar{\xi}^{ij})^T },
\end{equation}
where $i$ and $j$ indicate the tomographic redshift bins, $\xi_k$ is the data vector of the $k$-th realization, $\bar{\xi}$ is the mean data vector across all simulated realizations, and $\nu = N_s -1$ is the number of degrees of freedom given that the mean is estimated from the data.
If the number of data-points, $N_d$,  exceeds the number of realizations, $N_s$, we can neither ensure that the data covariance matrix is positive definite nor control the error in the data covariance matrix and its inverse. Therefore, we rebinned the data vector to reduce the number of points from 2310 to 770 by combining the angular bins in groups of three. Our final $\theta$ binning was chosen such that $N_s = 932$, $N_d = 770$ and hence $N_d<N_s$. 

\begin{figure}
\includegraphics[width=\columnwidth]{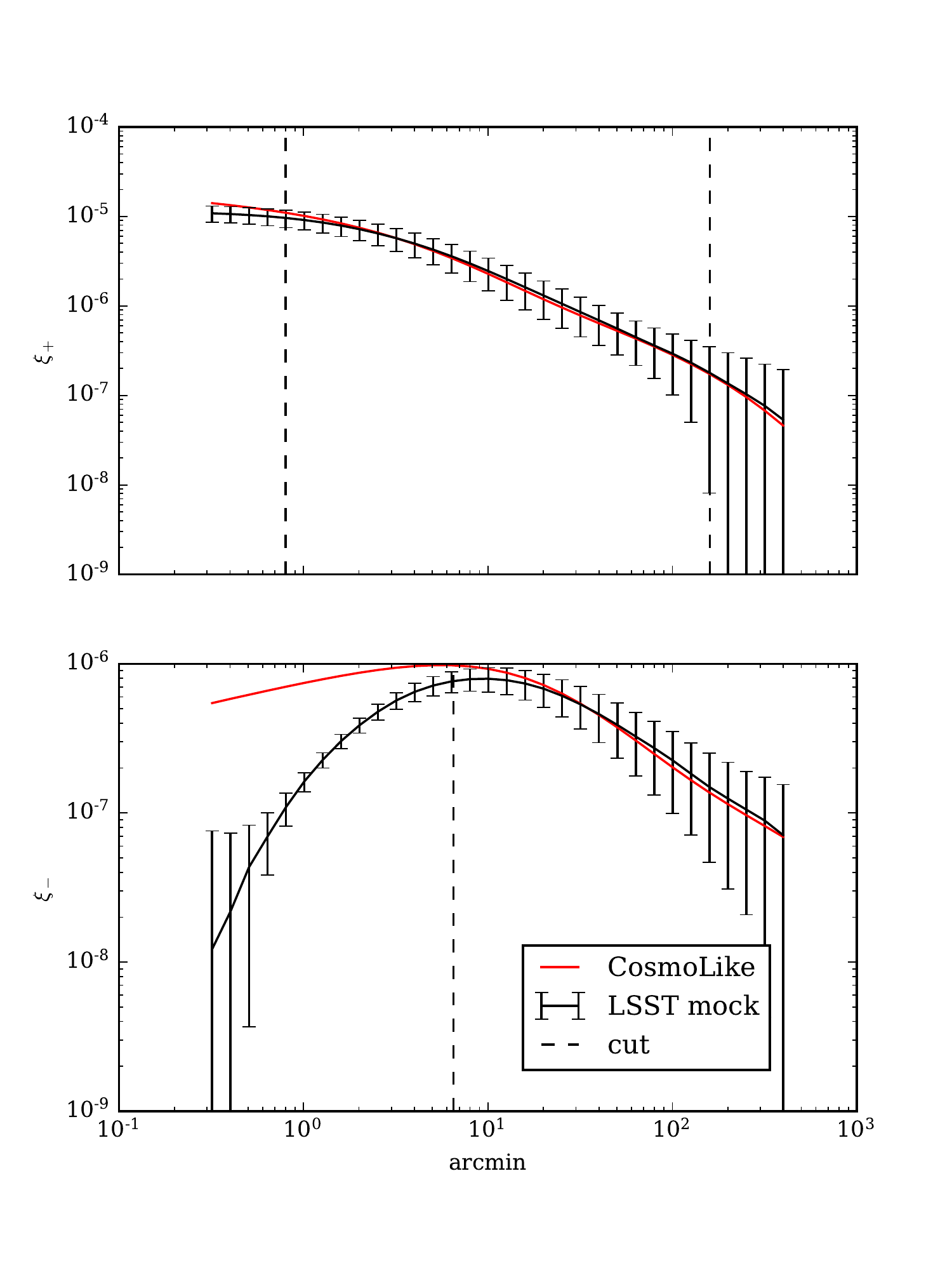}
\caption{\label{fig:xi_z1z1} Correlation functions $\xi_+$ (top) and $\xi_-$ (bottom) for the auto-correlation of the first tomographic bin ($z_1$, $z_2$) = (1, 1) of our mocks with shape noise and theoretical predictions from \textsc{CosmoLike}. The solid black curve shows the average and the 16\% and 84\% percentiles of the 932 mock realizations due to cosmic variance only. The adopted $\theta$ ranges for $\xi_\pm$ are shown by the vertical dashed lines, which were chosen due to limitations in the simulations as described in the text. The mismatch between the two curves is captured by the ratio ${\xi^{ij}_\text{theory}(\theta)}/{\left  < \xi^{ij}_\text{mock}(\theta) \right >}$.
} 
\end{figure}

An illustration of the data vector for a particular set of tomographic bins, the diagonal covariance matrix elements, and a comparison with analytic theory predictions is given in Fig.~\ref{fig:xi_z1z1}. The theoretical predictions are based on the HALOFIT method \citep{tsn12}, which models the nonlinear power spectrum with a fitting function. The HALOFIT model agrees well in general with the measurements from the simulations on the scales of interest, but there are still percent-level errors compared to the simulations \citep{2015MNRAS.450.2857H}. The finite box effect of the simulations also introduces power drops with respect to the theoretical predictions \citep{2015MNRAS.450.2857H}. The mismatch between the simulations and the theory will be compensated by a correction factor
\begin{equation}
\label{eq:rescaling}
{\xi^{ij}_\text{theory}(\theta)}/{\left  < \xi^{ij}_\text{mock}(\theta) \right >}.
\end{equation}
The correction factor above is not intended to fix these or other limitations of the simulations (which have some simplifications compared to reality, e.g., lack of baryonic physics). Rather, the correction is applied in order to assure that any bias we observe in our likelihood analysis comes from the covariance instead of the very small but nonzero intrinsic mismatch between the simulations and the theory.
Note that the correction factor is applied before the computation of the covariance, skewness and kurtosis, and is therefore included in these estimators.

\subsection{Modeling of observables}

For the simulated likelihood analyses in Sect.~\ref{sec:result} we employ the \textsc{CosmoLike} analysis and forecasting software package\footnote{www.cosmolike.info}.   
\textsc{CosmoLike} has been used in several forecasts exploring joint analyses of multiple cosmological probes \citep{eks14, kre17, ske17} and systematics mitigation strategies, such as the impact of baryons and intrinsic alignment \citep{2015MNRAS.454.2451E, 2016MNRAS.456..207K}. On the observational side, the code was used in a weak lensing analysis of Sloan Digital Sky Data \citep{heh14}, the analysis of DES science verification data \citep{btm16} and the recent DES Year 1 analysis \citep{2017arXiv170609359K,PhysRevD.98.043528,PhysRevD.98.043526}. 

We compute the linear power spectrum of the best fitting flat $\Lambda$CDM cosmology for WMAP9 + BAO + SN \citep{2013ApJS..208...19H} using the \cite{1999ApJ...511....5E} transfer function and model the non-linear evolution of the density field as described in \cite{tsn12}.
From the density power spectrum $P_\delta(k,z)$, we compute the shear power spectrum using the Limber and the flat sky approximations as 
\begin{equation}
\label{eq:pdeltatopkappa}
C_{\kappa \kappa}^{ij} (l) = \frac{9H_0^4 \om^2}{4c^4} \int_0^{\chi_\mr h} 
\mr d \chi \, \frac{g^{i}(\chi) g^{j}(\chi)}{a^2(\chi)} \pd \left(\frac{l}{f_K(\chi)},\chi \right) \,,
\end{equation}
with $l$ being the 2D wave vector perpendicular to the line of sight, $\chi$ denoting the comoving coordinate, $\chi_\mr h$ is the comoving distance to the horizon, $a(\chi)$ is the scale factor, and $f_K(\chi)$ the comoving angular diameter distance (throughout set to $\chi$ since we assume a flat Universe). The lens efficiency $g^{i}$ is defined as an integral over the redshift distribution of source galaxies $n(\chi(z))$ in the $i^\mr{th}$ tomographic interval
\begin{equation}
\label{eq:redshift_distri}
g^{i}(\chi) = \int_\chi^{\chi_{\mr h}} \mr d \chi' n^{i} (\chi') \frac{f_K (\chi'-\chi)}{f_K (\chi')} \,.
\end{equation}
We compute the cosmic shear two-point functions $\xi_\pm$ using the flat-sky approximation
\begin{equation}
\xi_{\pm}^{ij}(\theta) = \int \frac{dl\, l}{2\pi} J_{0/4}(l\theta) C_{\kappa \kappa}^{ij}(l)\,,
\label{eq:xi}
\end{equation}
with $J_n(x)$ the $n$-th order Bessel function of the first kind. 

\subsection{Likelihood functions and data covariance matrix}
From Bayes' theorem, we can compute the posterior distribution of the cosmological parameters. In the standard likelihood analysis, the likelihood function is parametrized as a multivariate Gaussian function

\begin{equation}
\label{eq:multivariate_gaussian}
L(\vec{\xi} \mid \vec{\pi}) \propto \exp[ -\frac{1}{2} (\vec{\xi} - \vec{\xi_\pi})^T \mathbf{C}^{-1} (\vec{\xi} - \vec{\xi_\pi})],
\end{equation}
where $\vec{\xi}$, $\vec{\pi}$ and $\mathbf{C}^{}$ denote the data vector, cosmological parameters and the covariance matrix respectively. The covariance matrix is fixed throughout the analysis.

Analytically computed covariance matrices are noise-free and can be factorized or inverted without complications. 
However, our covariance matrix estimated from numerical simulations is inherently noisy, and the noise level is affected by the number of realizations ($N_s =932$ in our case) and the size of the data vector ($N_d=770$).

To compute the likelihood function in the form of Eq.~\eqref{eq:multivariate_gaussian}, we need the inverse covariance matrix $\Psi$, also called precision matrix. An unbiased estimator of the precision matrix is given by \citep{2003Anderson,2007A&A...464..399H, 2013MNRAS.432.1928T}
\begin{equation}
\label{eq:psi}
\widehat{\Psi} = \frac{\nu - N_d -1}{\nu} (\hat{\mathbf{C}})^{-1}
\end{equation}
 in the case that the noise is Gaussian-distributed and the data points are statistically independent with $\nu = N_s -1$.

If the observed data is drawn from a multivariate Gaussian, \cite{2016MNRAS.456L.132S} show that after marginalizing over the noisy covariance matrix, the likelihood measured from the simulation realizations follows a multivariate t-distribution rather than a multivariate Gaussian.
An earlier correction proposed by \cite{2007A&A...464..399H} uses the unbiased inverse covariance matrix and keeps the multivariate Gaussian distribution to construct the likelihood function when the underlying distribution is Gaussian and the covariance is noisy. Thus, the Hartlap's correction method leads to
\begin{equation}
\label{eq:multivariate_gaussian_hartlap}
L(\vec{\xi} \mid \vec{\pi}) \propto \exp[ -\frac{1}{2} (\vec{\xi} - \vec{\xi_\pi})^T \widehat{\Psi} (\vec{\xi} - \vec{\xi_\pi})].
\end{equation}

The Sellentin-Heavens likelihood shows improvement over Hartlap's correction in terms of parameter inference with the marginalization over noise in covariance matrices. But it is still unclear how to extend the t-distributed likelihood function to cases where the underlying distribution of the data is non-Gaussian. In this paper, we adopt Eq. (\ref{eq:psi}) instead, since it is easier to apply to near-Gaussian distributions.
Since we quantify the difference between the Gaussian likelihood and the non-Gaussian likelihood through bias in cosmological parameter space, the bias depends more on the asymmetry of the likelihood distributions and thus is less sensitive to the difference between these two methods.

\subsection{PCA transformation}

For the shear correlation functions $\xi_\pm$, the multivariate Gaussian likelihood function with the form described in Eq.~\eqref{eq:multivariate_gaussian} is the most commonly used likelihood model in the literature. 
Since it is not trivial to build robust multivariate non-Gaussian likelihood functions, we perform the principal component analysis (PCA) transformations first on the data vector to remove the correlation between the data points.
PCA is an orthogonal transformation that transforms data points into coordinates without the linear correlations. It can be described as: 
\begin{equation}
\label{eq:pca_Lambda}
\hat{\Lambda} = Q^T \hat{\mathbf{C}} Q
\end{equation}
with columns of the transformation matrix $Q$ containing the eigenvectors of the covariance matrix $\hat{\mathbf{C}}$ estimated from the simulations. After the PCA transformation, the matrix $\hat{\Lambda}$ is diagonal and the diagonal elements are the eigenvalues of the covariance matrix $\hat{\mathbf{C}}$. The new coordinates are usually referred to as principal components, and the components are sorted according to decreasing variance. In addition, PCA automatically concentrates the information into a smaller number of principal components. For instance, \cite{2013MNRAS.431.3349H} studied the noisy covariance matrix of the matter power spectrum and found that the information content of the leading principal components remains stable as they raised the noise level by reducing the number of measurements from 200 to only 4. 

\begin{figure}
\includegraphics[width=\columnwidth]{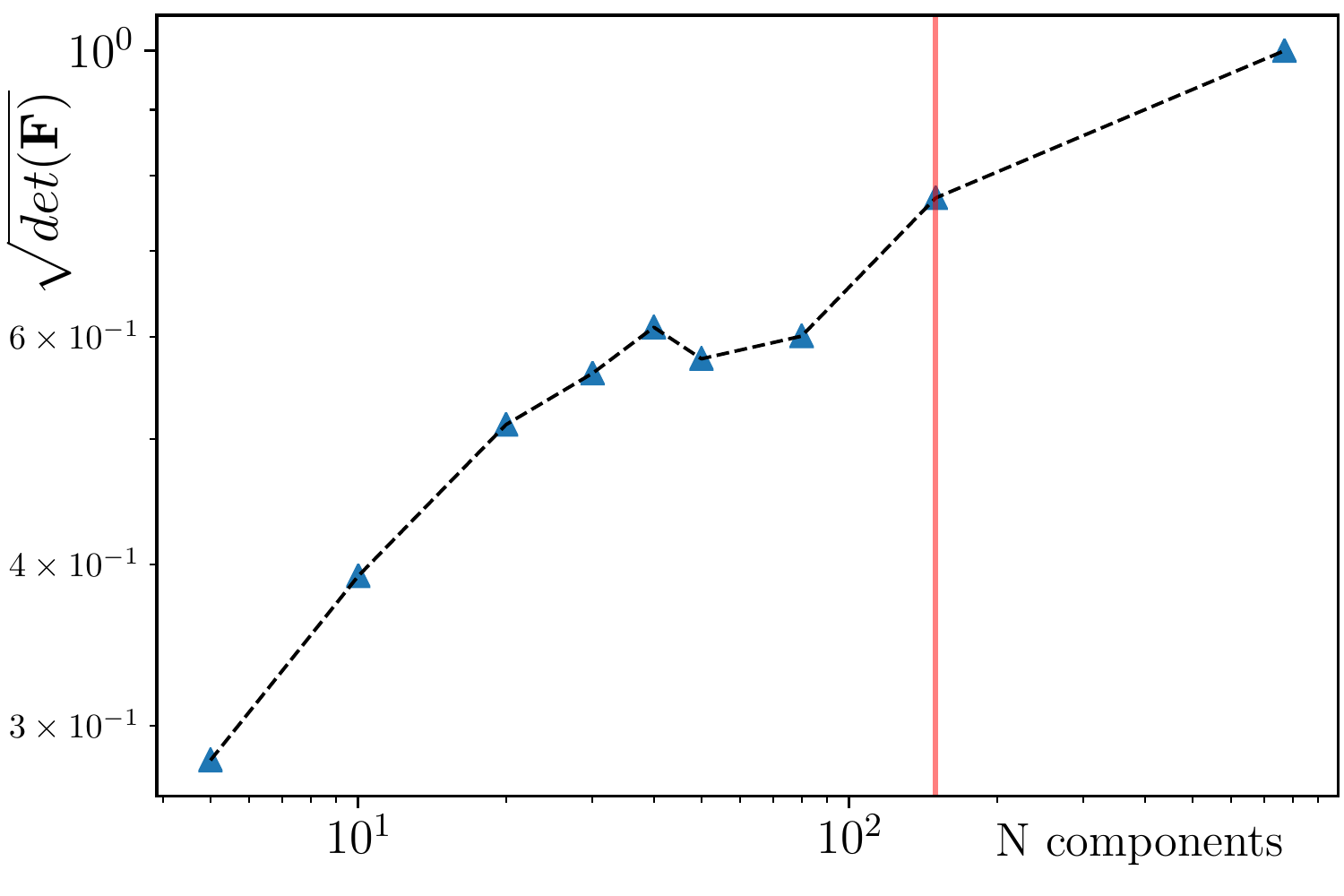}
\caption{\label{fig:det_inv_sqrt_noise} Information content of the training data versus number of principal components used in data analysis: the square root of the determinant of the inverse parameter covariance ($\Omega_{\rm m}$ and $\sigma_{\rm 8}$ only), i.e. the Fisher information matrix, quantifies the amount of retained information about the cosmological parameters. In the plot, the metric is normalized by the total information content (770 components). By keeping only 20\% of the data points (the red vertical line), we lose 23\% of information.
}
\end{figure}

Note that in Eq.~\eqref{eq:pca_Lambda} we use the covariance instead of its inverse in the PCA. It may sound counterintuitive that the data points with the highest variance, i.e.\ the highest uncertainty, contain more information. However, the principal components with highest variance are also those with the largest signal-to-noise ratio in our data (this was also found in \cite{2013MNRAS.431.3349H}), and thus contain more information. To quantify the information content on the training data contained in the first N~components, we perform a Markov Chain Monte Carlo (MCMC) analysis on the mock data with different numbers of principal components, and define the information content as the square root of the determinant of the inverse parameter covariance (Fisher matrix) of $\Omega_{\rm m}$ and $\sigma_{\rm 8}$, $\sqrt{det(\mathbf{F})}$  . Other parameters are fixed throughout the MCMC and a flat prior that limits parameters in the ranges 0.05<$\Omega_{\rm m}$<0.6 and 0.5<$\sigma_{\rm 8}$<1.1 is assumed.  
Figure~\ref{fig:det_inv_sqrt_noise} demonstrates the relation between the information content retained and the number of principal components used. It shows that if we apply data compression such that we have 80\% fewer data points, we lose only 23\% of the information content.

\subsection{Multivariate likelihood models}

After the PCA transformation, the data points are linearly uncorrelated. We next continue to use parametric and non-parametric functions to describe the likelihood distributions in the PCA space. 

In this paper we consider the following likelihood models for the PCA coordinates:

\begin{itemize}
\item{Gaussian function}
\item{non-Gaussian Edgeworth function}
\item{$k$-nearest neighbors}
\item{spectral series}
\end{itemize}

Here we describe the two parametric and the two non-parametric models listed above. The performance and results of the models are covered in more details in Sects. \ref{sec:model_assessment} and \ref{sec:result}.

Under the assumption that the underlying likelihood function measured in simulations is close to a multivariate Gaussian function, we approximate the multivariate likelihood function in the PCA coordinates as a product of parametric ``marginal distribution functions''. The independence of PCA components is a strong assumption, but it is the assumption that the standard multivariate Gaussian likelihood makes. The product of one-dimensional Gaussian distributions of PCA components is identical to the multivariate Gaussian likelihood. 

The Edgeworth function is a Gaussian function multiplied by correction terms constructed by its cumulants. It serves as an improvement upon the Gaussian likelihood. In this paper, we adopt Petrov's formula of the Edgeworth expansion \citep{1998A&AS..130..193B, 1962Petrov} and the coefficients in the expansion are fixed by the cumulants of the simulation data. In the case where the standard deviation $\sigma$=1, the first four terms in the expansion are  
\begin{equation}
\label{eq:edgeworth}
\begin{split}
\mathrm{Edgeworth}(x) & = \frac{1}{\sqrt{2\pi}} \exp \left[ - \frac{(\Delta x)^2}{2}\right] \\ 
& \cdot [ 1 
+ \frac{\kappa_3}{3!}He_3\left( \Delta x \right)
+  \frac{\kappa_4}{4!}He_4\left( \Delta x \right) \\
& +  \frac{10\kappa_3^2}{6!}He_6\left( \Delta x \right) 
+ \cdots
],
\end{split}
\end{equation}

where $He_n(x)$ are Hermite polynomials given by 
\begin{equation}
    \label{eq:hermite}
    He_n (x) = (-1)^n e^{x^2 /2}\frac{d^n}{dx^n}e^{-x^2/2},
\end{equation}
$\Delta x =  x-\mu$ and  $\kappa_n$ are cumulants. Moments and cumulants are two different sets of quantities that can summarize a distribution. Cumulants arise naturally from Fourier transformation. In the Fourier transformation, the probability density function $f(x)$ is transformed into 
\begin{equation}
\label{eq:cumulant_fourier}
\tilde{f}(k) = \int _{-\infty}^{\infty} e^{\mathrm{i}kx} f(x) \,\mathrm{d}x.
\end{equation}

The cumulants are then defined as the coefficients of the power series expansion 
\begin{equation}
\label{eq:cumulant_expansion}
\ln{\tilde{f}(k)} = \sum _{n=1}^{\infty} \kappa_n \frac{ (\mathrm{i}k)^n}{n!}.
\end{equation}

\begin{figure}
\includegraphics[width=\columnwidth]{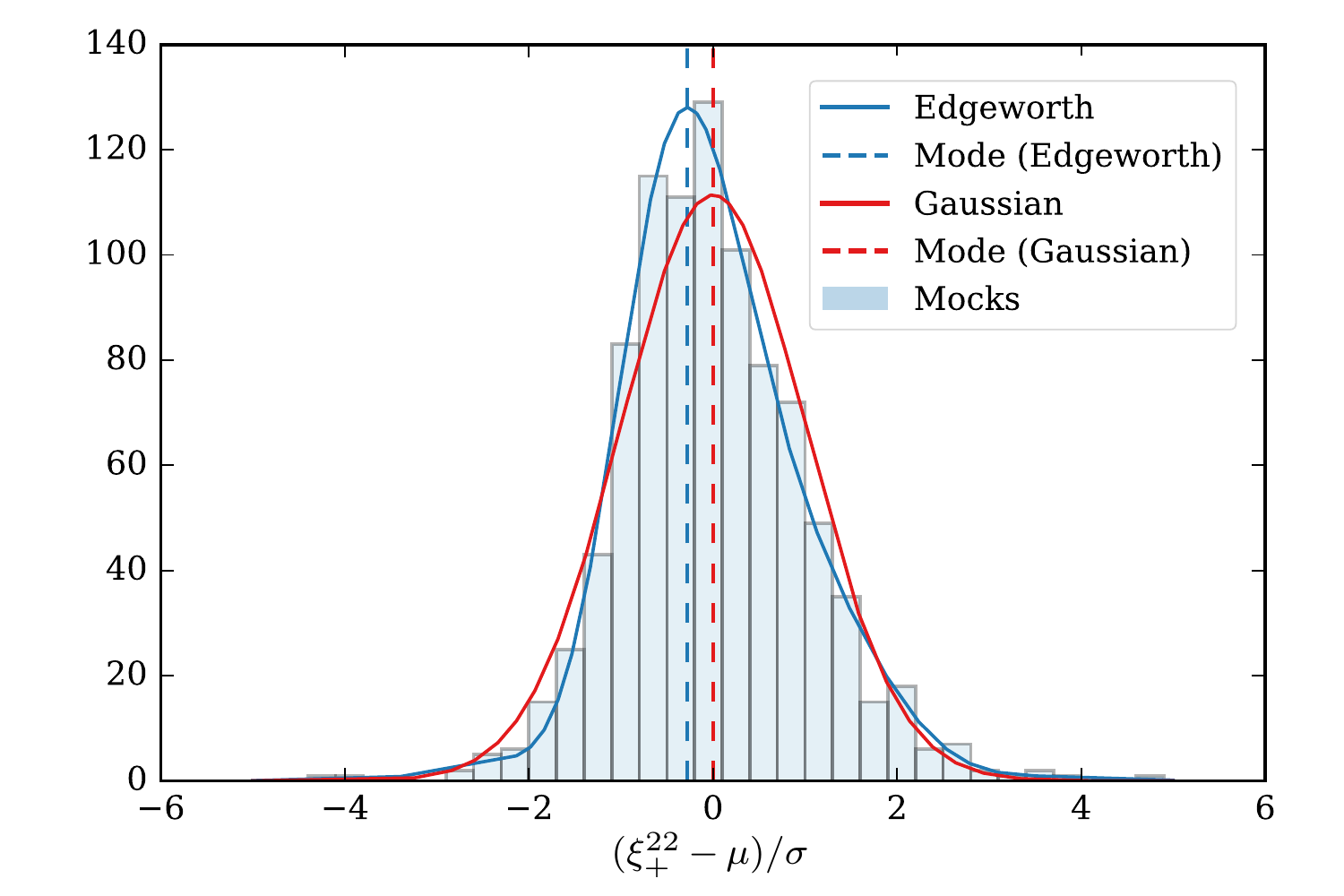}
\caption{\label{fig:1d_ex} Example of the non-Gaussian distribution of $\xi_{+}^{22}$ at $\theta$ = 159~arcmin in the mock weak lensing data with shape noise. This figure exhibits the low-level non-Gaussianity, and in particular the nonzero skewness, in the mock weak lensing data. If the distribution is skewed, the mean of the distribution deviates from the peak of the distribution (mode), which could lead to parameter biases if this feature of the likelihood is not adequately modeled.}
\end{figure}

Since the Edgeworth function is not guaranteed to be positive and could have oscillatory behavior, one should be careful with anomalies (negative probabilities) and avoid its use in strongly non-Gaussian cases. In our case, we do not find negative possibilities when modeling the marginal distributions of PCs with the Edgeworth expansion.
We show in Fig.~\ref{fig:1d_ex} an example of the marginal one-dimensional distributions of $\xi_+$ and the two parametric models. 

In addition to these parametric models, we also model the likelihoods more generally by estimating the high-dimensional density ratio $\beta (x) = f(x) / g(x)$ non-parametrically. Unlike the parametric methods that approximate the multivariate distributions as products of marginal distributions, our non-parametric methods do not assume independence.  In our case, we take $g(x)$ to be the Gaussian model. Once $\beta(x)$ is fitted, we can sample from the estimated density $f(x)$ by importance sampling with $g(x)$ as the proposal distribution. In this paper, the density ratio is estimated by non-parametric methods based on the k-nearest-neighbors kernel density estimator~\citep{Lincheng1985} and the Spectral Series estimators.

The k-nearest-neighbors estimator (knn) approximates the density at a point by a kernel smoothing applied to the k nearest neighbors of that point, and the Spectral Series estimator~\citep{izbickiLeeSchafer} combines orthogonal series expansion and adaptively chosen bases to construct non-parametric likelihood functions. Let $\{\psi_i\}$ be an orthonormal basis with respect to the data distribution; then the density estimator for spectral analysis has the form 
\begin{equation}
    \beta(x) = \sum{\alpha_i \psi_i(x)}.
\end{equation}
In low-dimensional non-parametric curve estimation, the basis is fixed to the usual choices, such as a Fourier basis. In the Spectral Series method, the basis is driven by data so as to capture the intrinsic dimensionality of the data~\citep{izbickiLeeSchafer}. 
With the density ratio, we are improving the Gaussian likelihood model using non-parametric methods. This is different from measuring the high-dimensional density with non-parametric models directly. Since the non-parametric models have more degrees of freedom compared to parametric models and make no assumption on the likelihood distributions, including the non-parametric methods make our list of models more complete.

\subsection{Skewness and Kurtosis}
\label{sec:skewness}
We examine particular departures from Gaussianity of the shear two-point correlation function by calculating higher moments of the distributions: skewness and kurtosis. The skewness can be quantified as the normalized expectation value of the third central moment
\begin{equation}
\label{eq:skewness}
\text{Skew}[X] = \frac{\mathbb{E} \left[(X-\mu)^3 \right]}{\sigma^3} 
\end{equation}
and it measures the asymmetry of the distribution. Increasing the number of samples $N_s$ does not reduce the skewness, but rather reduces the uncertainty of the skewness estimate, so repeated measurements would not remove the non-Gaussianity. Gaussian functions are symmetric; their skewness is zero. The distribution of shear correlation functions, however, is not perfectly symmetric. In Appendix~\ref{sec:third_moment} we derive for Gaussian fields the general expressions for the third moment of the likelihood of the shear correlation functions, from which the skewness can be predicted. Following the same derivation in Appendix~\ref{sec:third_moment}, the $n^{th}$ moment scales as $1/f_{sky}^{n-1}$ in general. 
Besides cosmic variance, the effect of shape noise can also be included in this framework.
 {We show that for the scales much smaller than the survey size, the third moment decreases with the survey size as ${f_\text{sky}^{-2}}$ and hence the skewness as defined in Eq.~\eqref{eq:skewness} decreases as ${f_\text{sky}}^{-1/2}$ ($\sigma\propto{f_\text{sky}}^{-1/2}$}).  As the scale $\theta$ approaches the survey window size, the third moment rises faster than $\sigma^3$ and thus the skewness will increase. 
 This trend is consistent with expectations from the Central Limit Theorem, which explains a decreasing skewness from an increase in survey area through the fact that the number of modes that are averaged over within given bin increases.

Besides the skewness, the asymmetry in the distribution is also captured by the mean-mode difference. In Appendix~\ref{mean_mode}, we limit the possible range of the mean-mode difference by assuming a unimodal distribution. For larger surveys, the mean-mode difference in terms of $\sigma$ also follows the same scaling relation as the skewness, $(\tilde \xi -\overline \xi)/\sigma \propto {f_\text{sky}}^{-1/2}$.

Additionally, we measure the kurtosis of the likelihood function as a metric for the level of non-Gaussianity. The kurtosis is defined as: 
\begin{equation}
\label{eq:kurtosis}
\text{Kurt}[X] = \frac{\mathbb{E} \left[(X-\mu)^4\right]}{\sigma^4} -3
\end{equation}
measures the symmetric outliers of the distribution. Since the fourth moment of the standard normal distribution equals 3, the kurtosis (or more precisely, the excess kurtosis) is defined as the normalized fourth moment minus 3. 

\section{Likelihood model assessment}
\label{sec:model_assessment}

The goal of this section is to introduce the statistical tools that we use to construct and assess the one-dimensional and multidimensional likelihood models in PCA coordinates. 
To properly compare different likelihood models and avoid overfitting, we use 10-fold cross-validation.
That is, the 932 realizations are partitioned into 10 non-overlapping subsets. For each experiment, 9 of the subsets are used to compute the mean and covariance of the Gaussian model or used to train the non-parametric models, while the remaining subset is used for testing. After the models are built, we draw samples from the models (90 realizations for each sample) and
compare these samples to the test samples using different two-sample
tests and distance metrics. This process is then repeated 10 times with each of the 10 data sets used once as the testing data.

In order to quantify the performance of the one-dimensional models for the likelihood of the data in PCA space, we perform cross-validation as described above with univariate Kolmogorov-Smirnov (KS) tests. The two-sample KS test statistic is the maximum distance between two empirical cumulative distribution functions. 
The p-value of the two-sample KS test statistic is defined under the null hypothesis that the two samples
are drawn from the same distribution. We take p<0.05 of this KS statistic as an indication that the null hypothesis is rejected and that the two sets of samples are not likely to have been drawn from the same distribution.

Besides tests on univariate distributions, we assess how close samples from the multivariate
models are to test samples, using two (non-parametric) multivariate
test statistics: the maximum mean discrepancy
(MMD;~\citealt{gretton2012kernel}) and the energy distance
(ED;~\citealt{szekely2004testing,baringhaus2004new}). 
The energy distance is a statistical distance between two probability distributions. It is defined as the square root of 
\begin{equation}
D^2(p,q) = 2 \mathbb{E} \|X-Y \| -  \mathbb{E} \|X-X' \| -  \mathbb{E}\|Y-Y'\|
\end{equation}
where $\mathbb{E}$ is the expectation value, $X$, $Y$, $X'$, and $Y'$ are independent random vectors, the distribution of $X$ and $X'$ is $p$, and the distribution of $Y$ and $Y'$ is $q$. Here we use the Euclidean metric (and a sample estimate of the above expression). The  maximum mean discrepancy can then be seen as a generalization of the energy distance 
to reproducing kernel Hilbert spaces. More specifically, we use a Gaussian kernel  $ K_h(x,y)$ with bandwidth $h$ to measure the similarity between two vectors $x$ and $y$, and we define our MMD test statistic as the MMD sample estimate:
\begin{equation}
\begin{split}
T & = \frac{1}{m^2} \sum_{i=1}^{m}\sum_{j=1}^{n} K_h(X_i,X_j)
-\frac{2}{mn} \sum_{i=1}^{m}\sum_{j=1}^{n} K_h(X_i,Y_j) \\
& +\frac{1}{n^2} \sum_{i=1}^{n}\sum_{j=1}^{n} K_h(Y_i,Y_j).
\end{split}
\end{equation}
where $X_1,\ldots,X_m \sim p$ and $Y_1,\ldots,Y_n \sim q$. 

The two test statistics above provide quantitative distance measures between samples drawn from two sampling distributions of weak lensing correlations functions $p$ and $q$. In the case of weak lensing analysis, $X$ and $Y$ in the above notations are multivariate shear correlation function data vectors drawn from the hold-out simulation data or the trained models. These test statistics or distance metrics\footnote{We use the publicly available R packages
  ``kernlab'' and ``energy''. We implement MMD with the Gaussian
  kernel and determine the bandwidth by the median heuristic method, i.e. $h^2 = \text{median}(\lVert X_i - X_j \rVert ^2)$. } are
invariant to orthogonal transformations (such as a PCA rotation) and
often used for comparing higher-dimensional data.

\begin{figure*}
\includegraphics[width=14cm]{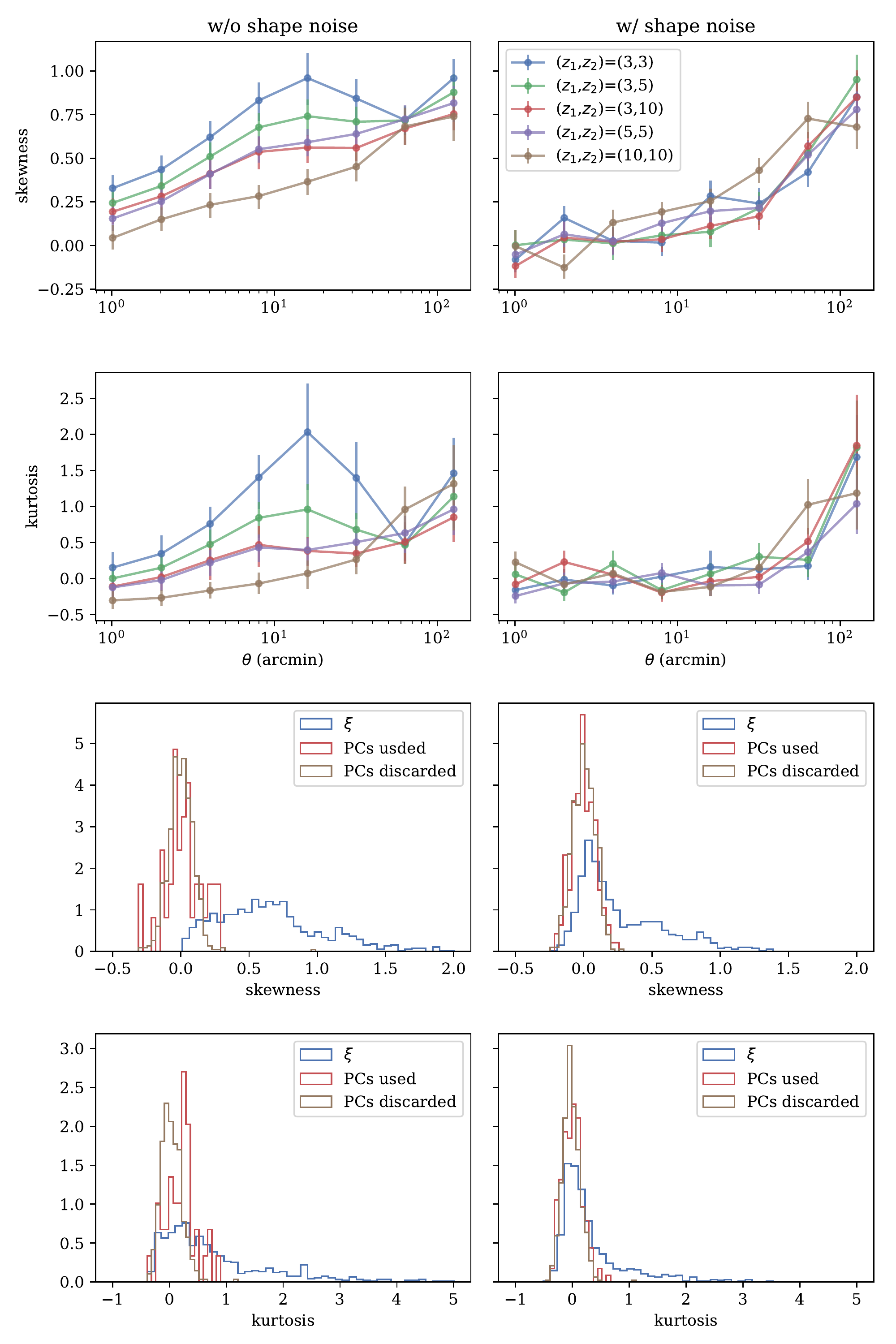}
\centering
\caption{\label{fig:sk+kur} The top two rows show the skewness and kurtosis of $\xi_+$ in tomographic bins ($z_1$, $z_2$)=(3,3), (3,5), (3,10),  (5,5) and (10,10) for data without shape noise (left column) and data with shape noise (right column). The curves for data without shape noise exhibit strong non-zero skewness and kurtosis and demonstrate particular departures from Gaussianity in the marginal distributions.  At lower redshift, the skewness and kurtosis are more significant.  The non-Gaussianity is not as strongly detected in data with shape noise. That is because the mock data is dominated by shape noise in most of the scales that we consider. Only at scales around 100~arcmin does the data with shape noise start to show comparable non-Gaussianity as data without shape noise. The bottom two rows are normalized histograms of skewness and kurtosis of 1-D distributions of the 770 data points before and after PCA transformation. We divide the principal components into the ones we use (40 components for noise-free data and 150 components for noisy data) and the ones we discard. The level of skewness and kurtosis of the principal components used for our analysis is below that for the $\xi_\pm$ values, and is distributed around zero. }
\end{figure*}

\begin{figure*}
\includegraphics[width=\textwidth]{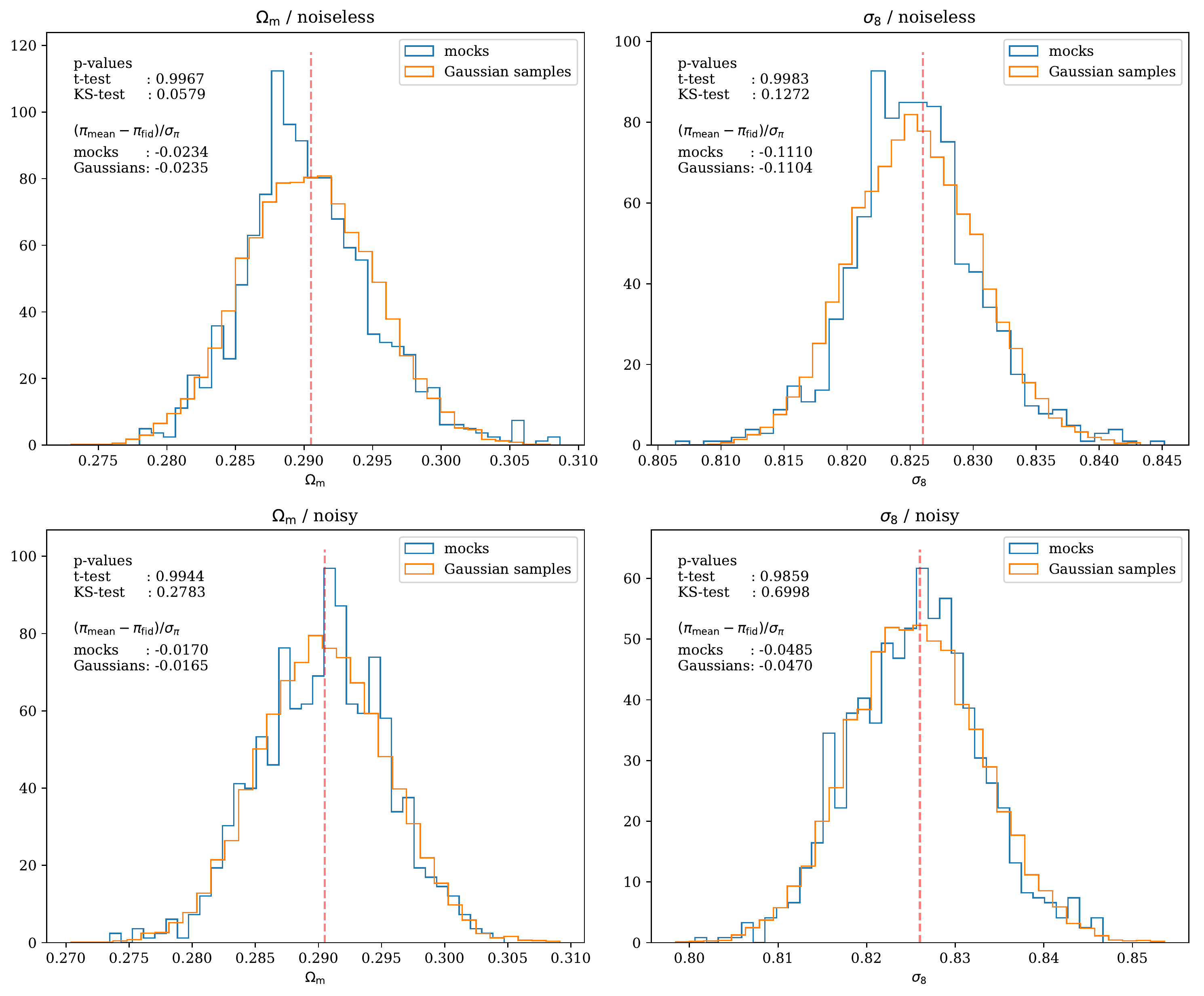}
\caption{\label{fig:maxlike_1d} One-dimensional maximum likelihood estimates of $\Omega_{\rm m}$ and $\sigma_{\rm 8}$ for data without shape noise (upper panels) and data with shape noise (lower panels). Besides the estimates based on the 932 realizations of correlation functions in the different simulation realizations (blue), the line in orange shows the parameter estimates of ten thousand Gaussian-distributed samples with the same mean and covariance matrix as the simulation realizations. By comparing the two histograms, we can estimate the impact of the Gaussian likelihood assumption with respect to parameter inference. The p-values of two-sample statistics, including the independent T-test and the KS-test, and the deviations of the mean values from the true point (red dotted line) are listed at the upper left corner of each panel. None of the statistics is below the threshold p-value of 5\%, meaning that, for all cases, there is no significant difference between the two distributions and that there is no significant parameter bias due to the Gaussian likelihood assumption.
}
\end{figure*}
\section{Results}
\label{sec:result}

\subsection{Biases due to non-Gaussian distributions of $\xi_\pm$}\label{subsec:bias-xi}

In the previous sections, we introduced the approach of building and assessing the likelihood distributions. Here we apply the tools to examine the biases due to non-Gaussianity in distributions of the shear correlation functions $\xi_\pm$.

If the likelihood function of $\xi_\pm$ is a multivariate Gaussian function, then its marginal distributions are Gaussian by construction. However, we detect non-Gaussianity in the marginal distributions of $\xi_\pm$ for many values of $\theta$ and tomographic bins. 
The finding that the likelihood of $\xi_\pm$ is skewed was previously discussed in \cite{2018MNRAS.473.2355S}. They use CFHTLenS Clone simulations for the tomographic analysis of the CFHTLenS data that provide 1656 semi-independent simulations for the 210 data points of CFHTLenS. For a specific data point at $\theta = 35$ arcmin, they found that the most likely lensing amplitude is about 5\% below the mean, so the distribution is `left-skewed'.

In Fig.~\ref{fig:sk+kur} we show the non-zero skewness and kurtosis of the 1-D distributions of $\xi_\pm$ in selected tomographic bins, and of PCA coordinates, for data with and without shape noise. For the shape noise-free data, the skewness and kurtosis both decrease as redshift increases. The magnitude of non-Gaussianity shown in the skewness and kurtosis is statistically significant and peaks roughly at $\theta = 20$~arcmin. It is difficult to gain insight into the $\theta-$ dependence of the skewness of the shear 2PCF, since the latter is an integral over the $C(l)$ values with highly oscillating filter functions $J_{0/4}$. 

Most of the scales that we consider are dominated by the shape noise, which strongly suppresses the  skewness and kurtosis. At scales around 100~arcmin, the skewness and the kurtosis start to reach a comparable level as in the shape noise-free case, since the shape noise is relatively less important at large scales.
Note that these results for skewness and kurtosis of the marginal 1D likelihoods of $\xi_\pm$ values do not fully represent the level of non-Gaussianity in the multivariate observable space, since the $\xi_\pm$ values are highly correlated across $\theta$ and redshift bins.

Compared to $\xi_\pm$, the sampling distribution of the data in PCA space has far less significant skewness and kurtosis. The PCA components are linear combinations of $\xi_\pm$ values. During the transformation, a large number of the $\xi_\pm$ data points are added with positive and negative weights. As a result the linear combination then has reduced skewness and is closer to Gaussian.
This is shown in the last two rows of Fig.~\ref{fig:sk+kur}, which compares the level of skewness and kurtosis in the marginal 1D distribution of $\xi$ and in PCA components. 
We again note that the multivariate sampling distribution is not affected by the PCA transformation, the marginal distributions can change due to rotations and the resulting marginal likelihoods depend on the specific forms of rotations. After the PCA rotation, we find that the 1-D likelihoods empirically become more Gaussian, as illustrated in the last two rows of Fig.~\ref{fig:sk+kur}.

To estimate the biases in cosmological parameter due to the non-Gaussianity in distributions of $\xi_\pm$, we apply the maximum likelihood method. Maximum likelihood estimation provides an intuitive way of estimating the biases in cosmological parameters due to a failure to model the non-Gaussian distributions of weak lensing two point functions. 
In this method, we compute the maximum likelihood estimate of cosmological parameters of each realization in the mocks with the Gaussian covariance. Given our 932 simulations, this yields 932 MLEs in cosmological parameter space. Limited by the number of realizations, we perform the maximum likelihood estimation for individual parameters, either $\Omega_{\rm m}$ or $\sigma_{\rm 8}$ (separately), with the other parameters fixed to their true values in the simulations. To avoid a bias due to a small mismatch between the mocks and the theoretical prediction (see Fig.~\ref{fig:xi_z1z1}), the elements of data vectors for each simulation realization are rescaled by  ratios ${\xi^{ij}_\text{theory-fid}(\theta)}/{\left  < \xi^{ij}_\text{mock}(\theta) \right >}$.

In order to estimate the impact of incorrect likelihood models, we also perform the maximum likelihood estimation on 10,000 Gaussian-distributed samples of two-point correlation functions. We first model the likelihood distributions of the simulation realizations with a multivariate Gaussian, and then draw samples from them. The samples share the same mean value and covariance matrix as the mocks but follow a multivariate Gaussian distribution. Hence the difference between the 1-D parameter estimates of the two sets of samples (mocks and Gaussian samples) comes solely from the incorrect assumption of likelihood functions.

We show in Fig.~\ref{fig:maxlike_1d} the 1-D parameter estimates of $\Omega_m$ and $\sigma_8$ for the mock data and the 10,000 Gaussian samples with and without shape noise. In finding the maximum likelihood estimates, only one parameter ($\Omega_{\rm m}$ or $\sigma_{\rm 8}$) is explored at a time. To compare the 1-D distributions, we consider two statistical tests: the independent T-test and the KS-test. The T-test determines whether there is a  significant difference between the mean values of two samples. Judging from the high p-values of the T-test for all the four cases in Fig.~\ref{fig:maxlike_1d}, we do not see significant shifts between the average values of the two sets of samples. This indicates that the mean values of 1-D parameter posteriors are not affected by the non-Gaussian marginal distributions of shear correlation functions that we observe in the mock data. In addition, we find that the mean values in the four panels are located around the fiducial values of parameters ($\Omega_{\rm m}=0.2905$ and $\sigma_{\rm 8}=0.826$). The deviations from the true value with respect to the uncertainties are listed in Fig.~\ref{fig:maxlike_1d}.

Besides the T-test, we also report the p-values of the two-sample KS-test. The p-values of the KS statistics are all above the 5\% threshold, meaning that we do not detect significant difference between the 1-D distributions of parameters of the two samples. With the T-test we learn that the average cosmological parameters are not shifted by the non-Gaussian features of likelihoods in the mocks. The KS-test results suggest that the 1-D distributions of parameters are neither skewed nor distorted in other ways significantly by failure to model the left-skewed marginal distributions of shear correlation functions.

\subsection{Modeling the distributions of principal components}
\label{sec:modeling}
In this subsection, we show results of attempts to model the likelihood function of PCA components.

\begin{figure*}
  \includegraphics[width=12cm]{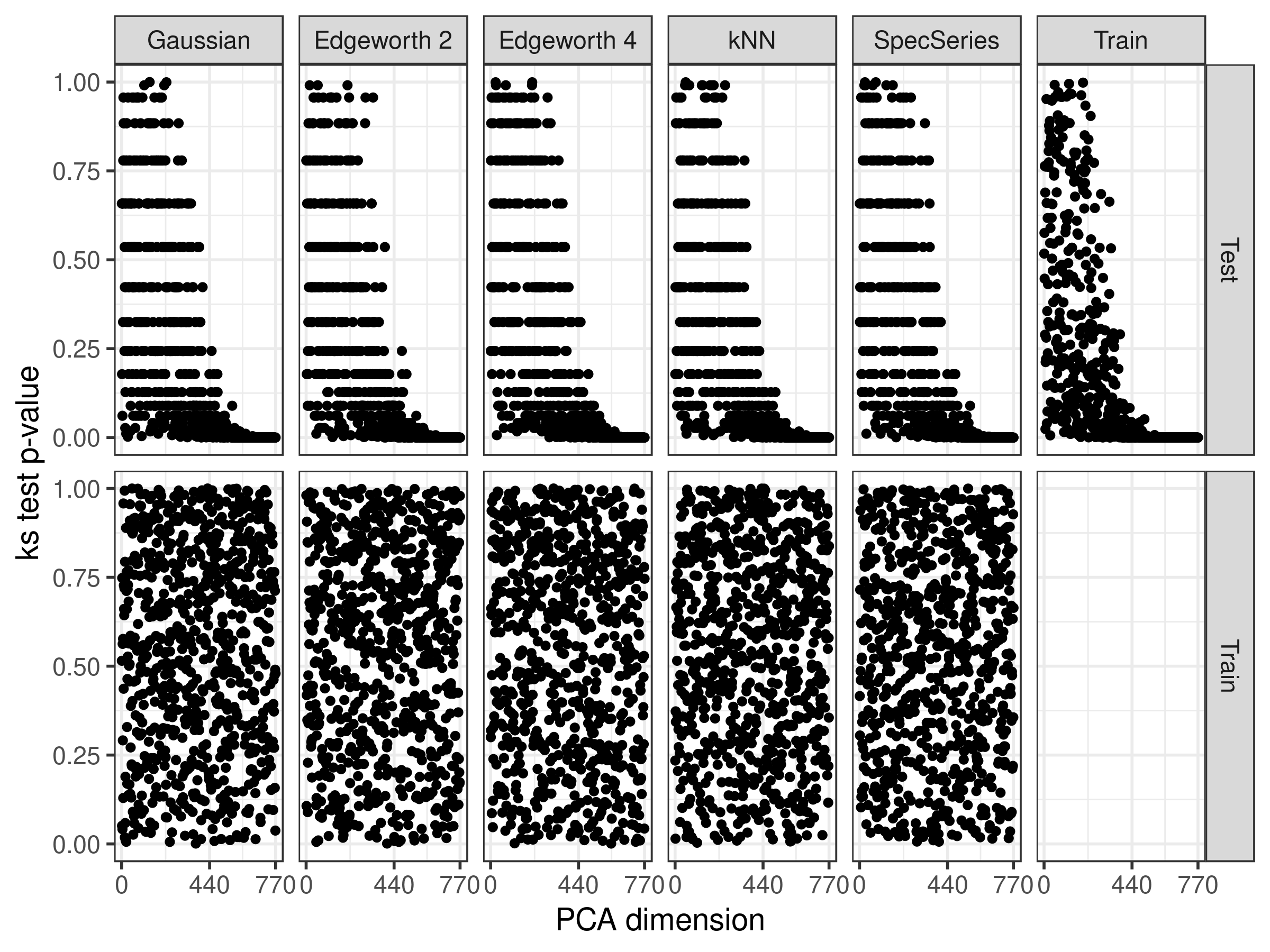}
  \caption{\label{fig:2sample_test} 
 KS-test statistics for each PCA dimension on testing and training data. The first five columns show the p-value against the data drawn from the trained models and the upper panel of the last column shows the p-value of training data against testing data. All models have uniformly distributed p-value for training data, showing good univariate fitting performance. The models fail to generalize to testing data for higher-order PCA dimensions. The coarseness of the points is due to the smaller number of data points in the test set. }
\end{figure*}

When we perform KS tests of the univariate Gaussian and Edgeworth models on the training data set, all of the PCA components have $p$-values that are approximately uniformly distributed between 0 and 1 as shown in the bottom row of Fig. \ref{fig:2sample_test}.
This means that both of the parametric models have good univariate fitting performance on the 90\% training data for data with and without shape noise.  
When we perform KS tests for these models against the testing data (top row of Fig. \ref{fig:2sample_test}), however, the $p$-values concentrate around 0 for higher principal components for all the parametric models we consider, including Gaussian, Edgeworth function to the second order and Edgeworth function to the fourth order. This indicates that neither one of the models generalizes well to hold-out data for the computed low-variance principal components as a result of the PCA decomposition. Including the components that are mostly dominated by noise would overfit the data. In addition, Fig. \ref{fig:2sample_test} shows that the Gaussian model fits decently to the leading principal components of the current data. 
We could not noticeably improve the fitting performance by using more complicated models such as the 1-D Edgeworth function, but we can avoid overfitting by adopting the PCA framework and discarding the high-order principal components with additional benefit of data compression.
Note that the KS test statistic is the largest difference between the two empirical CDFs. The KS-statistics take on discrete values, since the empirical CDFs are discrete due to the finite number of samples. 

An analysis of variances also shows that the first 40 components
capture over 99.9\% of the variance of the training data without intrinsic shape noise, whereas the
corresponding number of components for data with shape noise is 400. 

\begin{figure*}
  \includegraphics[width=12cm]{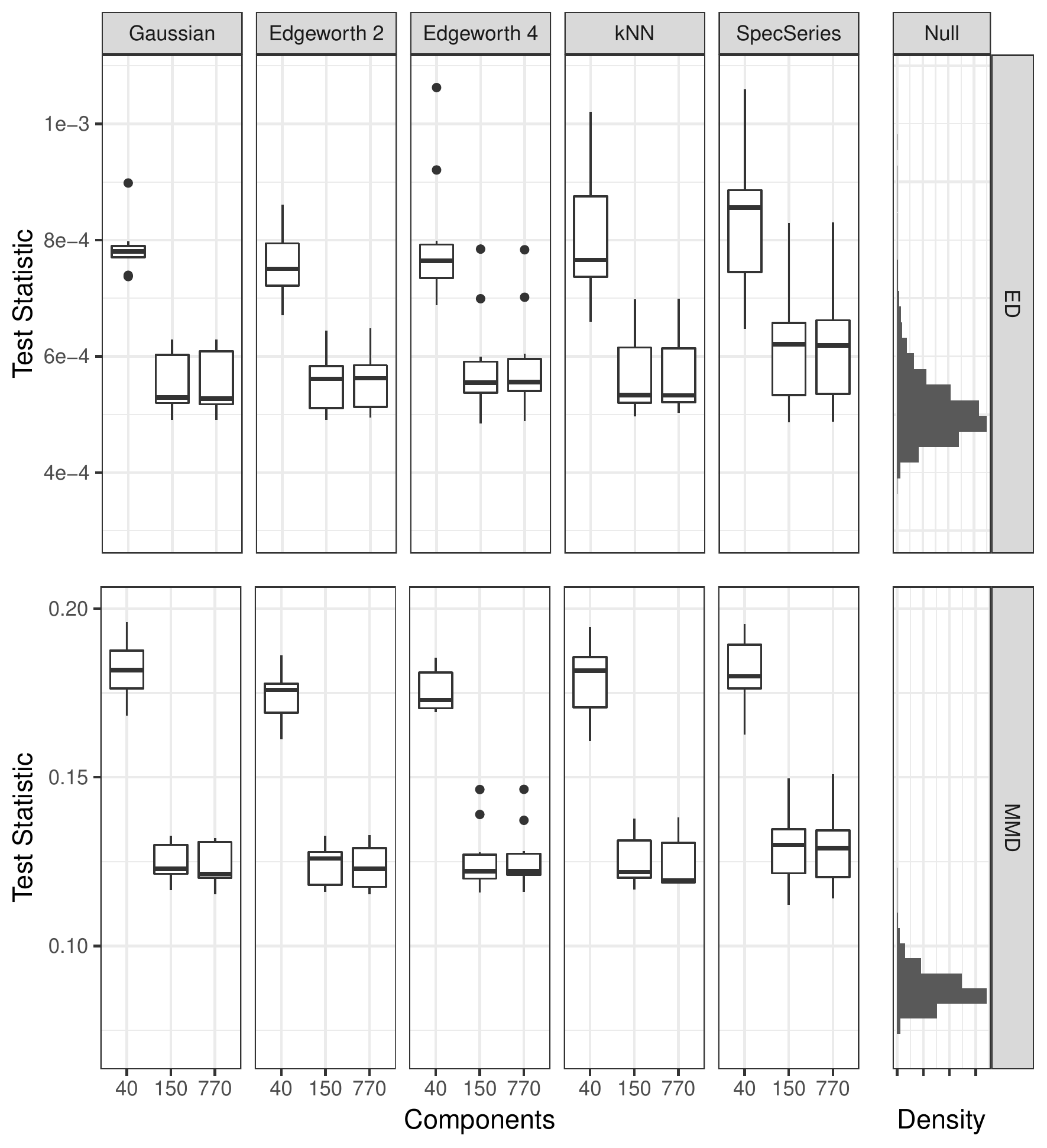}
  \caption{\label{fig:2sample_multi_test_with_noise} Box plots of test statistics for multivariate two-sample tests of different models on hold-out data (10 folds) with shape noise. The statistics are distance metrics of the two multivariate two sample tests (ED and MMD). Smaller statistics mean better agreement between the samples drawn from the trained model and the testing data set. All models are roughly equivalent in generalization performance, but we need to include many more dimensions (150 components are needed for data with shape noise) than in the noiseless case where 40 components are sufficient. The boxes show the range between the first and the third quartile, with the median labeled by the central bar in the boxes, the variability outside the upper and lower quartiles labeled by the vertical lines, and outliers as individual dots. The estimated null distribution is depicted in the final column; this represents the distribution of test statistics under the assumption that the estimated model is correct.  For MMD we see that our observed statistics are outside the range of the null distribution. This means that the MMD statistic is discriminatory enough for our data set. It also implies that improving the likelihood models is possible, but leveraging the advantages of more flexible models may require much larger amounts of data.
  }
\end{figure*}

Figure~\ref{fig:2sample_multi_test_with_noise} shows the multivariate
performance of different likelihood models including parametric models
(Gaussian, Edgeworth to the 2nd order, Edgeworth to the 4th order) and non-parametric models (k-nearest-neighbor
and Spectral Series) for data with shape noise. The lower
the statistics (that is, the smaller the distance between samples from
the model and samples from the test sets), the higher the performance. 
The last column of Fig.~\ref{fig:2sample_multi_test_with_noise} shows the null distribution of the test statistics, and tells us what distance values to expect if two samples or sets (with 93 realizations each) were drawn from the true distribution of the data. We estimated this distribution by repeatedly resampling the unmodeled data without replacement, and then in each fold computing distances between these samples and the hold-out sample. 

For the data without shape noise, models based on 40 components give
good performance on hold-out data, whereas we need around 150 components in the presence
of shape noise as seen in Fig.~\ref{fig:2sample_multi_test_with_noise} for similar performance. 
We do not see significant differences in performance
between the multivariate Gaussian model and the other parametric or non-parametric methods. The estimated null distribution in the last column of Fig. ~\ref{fig:2sample_multi_test_with_noise} indicates that the MMD statistic is powerful enough to discriminate between models for our data set. There is a significant difference betweeen the MMD statistics of models and the real data, implying that there is still some room for improvement. But improving the models would require an order-of-magnitude increase in the number of realizations, which is currently beyond our reach. Hence, our
conclusion is that given the current number of realizations and inherent data
noise, more complex multi-dimensional likelihood models do not seem to improve upon a
simple Gaussian likelihood model typical of an LSST-like survey.

\begin{figure}
  \includegraphics[width=\columnwidth]{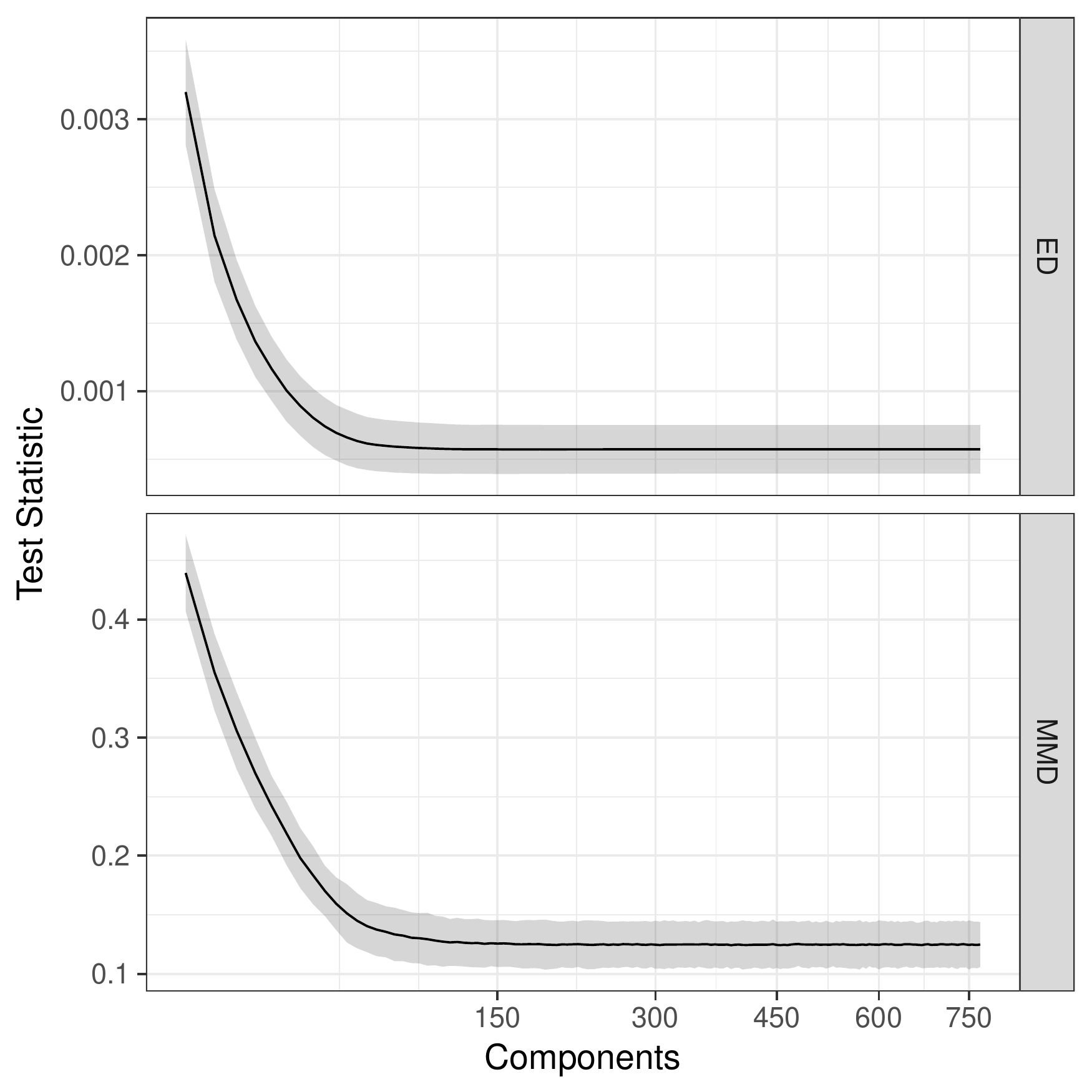}
  \caption{\label{fig:2sample_noise_curve} Test statistics versus number of
    components for multivariate two-sample tests on hold-out data with shape noise; the solid line shows the median value and the shaded area represents the interquartile range. Using 150 components gives similar generalization performance as using more components. }
\end{figure}

The number of components that retains $99.9\%$ of the variance is just an estimate. In order to find the optimal number of components to use, we apply our
multivariate testing framework to a $k$-dimensional Gaussian
likelihood model for different candidate values of $k$.
Figure~\ref{fig:2sample_noise_curve} indicates that we can further reduce the
dimension of the model to 150 for
data with shape noise without oversmoothing or ``underfitting'' the data. 
For data without shape noise, we can reduce the dimensionality to 40, the same as for our previous 99.9\% variance estimate.

\subsection{Biases due to non-Gaussian distributions of principal components}

In Sect.~\ref{subsec:bias-xi}, the maximum likelihood method was used to directly estimate cosmological parameter biases due to the non-Gaussian likelihood function of weak lensing shear correlation functions.  This method of estimating the level of parameter biases when using the Gaussian likelihood approximation suggested that those biases are subdominant to statistical uncertainties in the mocks. 
In this subsection, we consider the multivariate PCA models that we have built and assessed in Section~\ref{sec:modeling} and estimate parameter biases by comparing the MCMC chains of different models with \textsc{emcee} \footnote{https://github.com/dfm/emcee} \citep{2013PASP..125..306F} MCMC sampler. The MCMC chains only explore $\Omega_m$ and $\sigma_8$ with the other parameters fixed, and we use flat priors for $\Omega_m$ and $\sigma_8$ with range 0.05<$\Omega_{\rm m}$<0.6 and 0.5<$\sigma_{\rm 8}$<1.1. In the MCMC sampling, we assume a constant covariance estimated from the simulations at the fiducial cosmology, regardless of the changes in $\Omega_m$ and $\sigma_8$.

\begin{figure}
\includegraphics[width=\columnwidth]{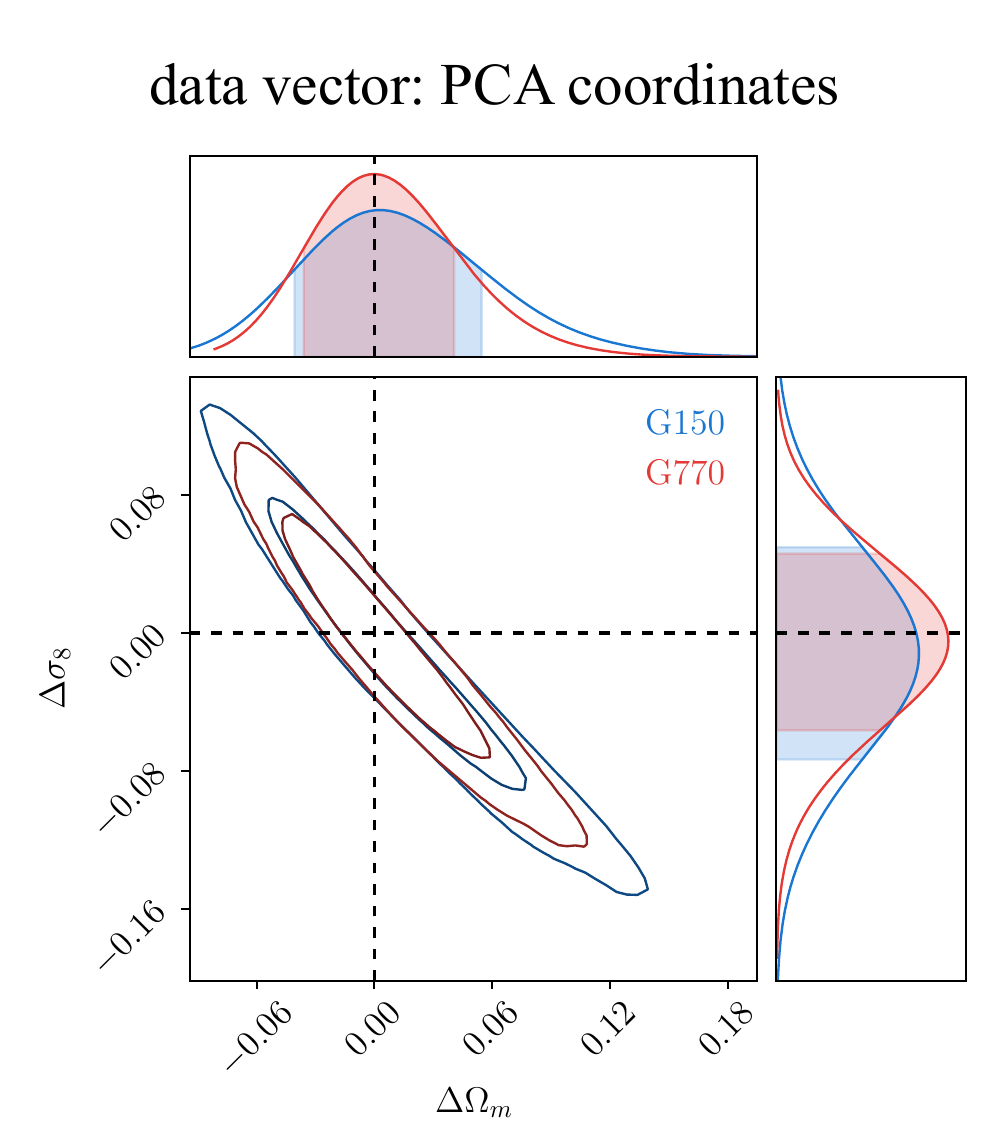}
\caption{\label{fig:MCMC_noise_compression} Parameter constraints for the Gaussian likelihood models with 150 and 770 principal components using mock data with shape noise. This figure compares the same Gaussian model with 150 and 770 principal components. There is a $\sim$20\% difference in the contour area between 150 and 770 components due to the loss of information when using fewer components. Note that the G770 curve also represents the standard multivariate Gaussian likelihood of $\xi_{\pm}.$
}
\end{figure}

\begin{figure}
\includegraphics[width=\columnwidth]{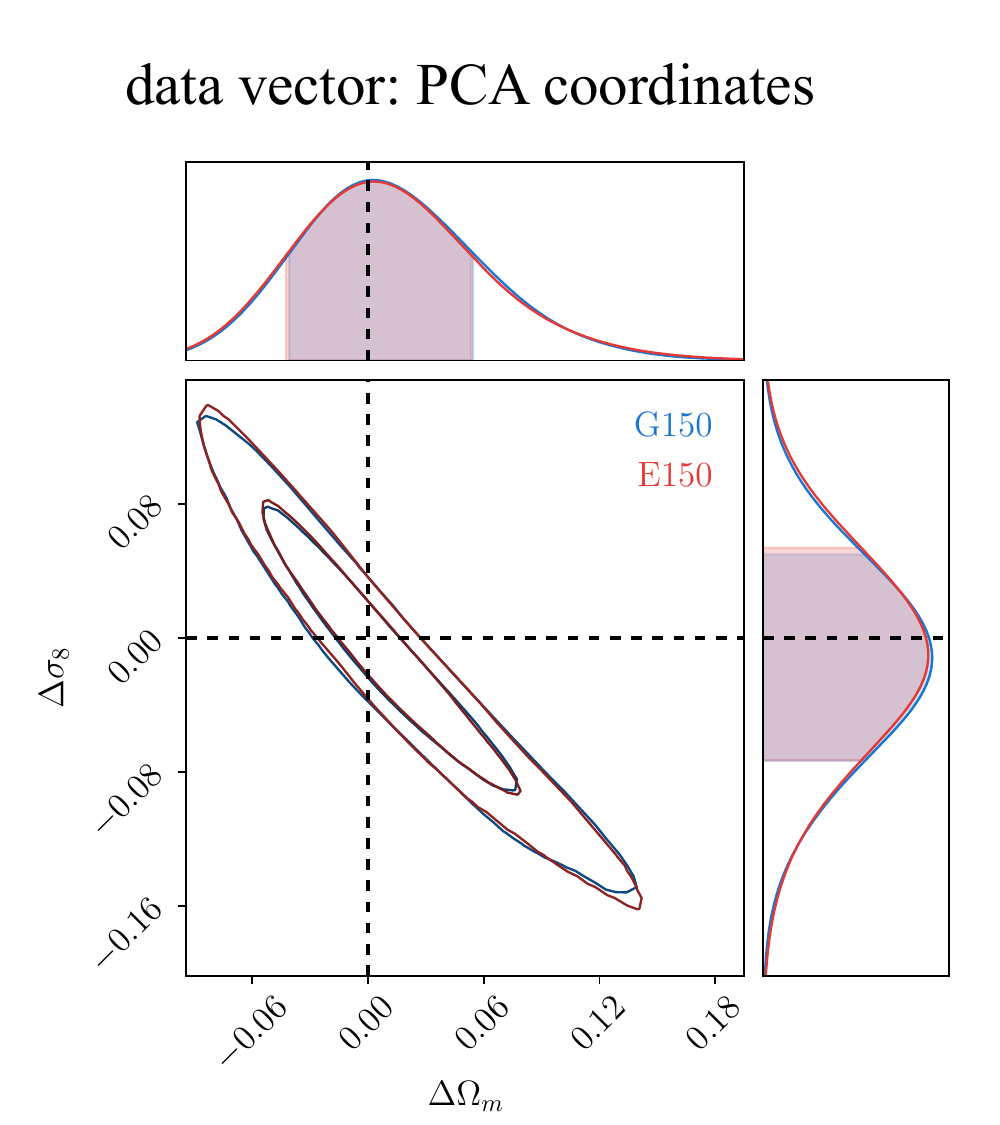}
\caption{\label{fig:MCMC_noise_GE} Parameter constraints for the Gaussian and the Edgeworth model with 150 components on the shape-noisy mock data. This figure compares the Gaussian model and the Edgeworth model with the same number of principal components (150 components). No significant difference is found between the posteriors of the cosmological parameters when using a Gaussian or Edgeworth model for the likelihood.}
\end{figure}

Figures~\ref{fig:MCMC_noise_compression} and \ref{fig:MCMC_noise_GE} show contour plots of posteriors for the cosmological parameter constraints derived from the mock data with shape noise. The posteriors are re-centered to the true value as a result of the rescaling by the ratio in Eq.~\eqref{eq:rescaling}.  Fig.~\ref{fig:MCMC_noise_compression} compares the performance of models with different numbers of principal components and shows the effect of data compression. The 20\% difference in the contour areas is due to information loss in data compression. Note that the model with all 770 PCA components and 1-D Gaussian distributions is strictly identical to the standard multivariate Gaussian likelihood of $\xi_\pm$. Hence the contour labeled as G770 in Figure~\ref{fig:MCMC_noise_compression} is also what we expect from the standard Gaussian likelihood analysis. Figure~\ref{fig:MCMC_noise_GE}, on the other hand, compares Gaussian and non-Gaussian Edgeworth models in the PCA coordinates to demonstrate any biases due to non-Gaussian likelihood functions. Consistent with the results of the maximum likelihood estimation, no significant difference is found between the contours of the Gaussian model and the Edgeworth model for the mock data with shape noise. 

\section{Conclusions and Discussions}
\label{sec:conclusion}

It is well known that the multivariate likelihood function of weak lensing shear correlation functions is not a Gaussian. Approximating the likelihood with a Gaussian should introduce associated biases in recovered cosmological parameters; however, whether this would cause significant biases for weak lensing shear correlation functions for upcoming lensing surveys and invalidate the mainstream multivariate Gaussian likelihood assumption has not yet been established. In this paper, we make further advances towards answering this question by modeling and testing non-Gaussian likelihood functions with simulated weak lensing data, and estimating the resulting biases on cosmological parameters.

We have measured the level of non-Gaussianity of shear correlation function likelihoods in weak lensing simulations and suggest avenues to improve corresponding cosmic shear likelihood analyses. The simulations are based on 100 deg$^2$ lines-of-sight with the same source redshift distribution and number density as expected for LSST. A systematic approach to constructing the likelihood models was used, and biases in the parameter space ($\Omega_m, \sigma_8$) due to the assumption of a multivariate Gaussian likelihood model were assessed. 

We explored the non-Gaussianity in univariate distributions of our data vectors and find detectable non-Gaussianity in the marginal 1D likelihoods of shear correlation functions in $\theta$ and redshift bins. 
Even though the marginal distributions of $\xi_\pm$ provide hints on the non-Gaussianity, the linear transformation of PCA reduces the level of skewness and kurtosis in the principal components significantly. 

Besides the tests on marginal 1D likelihoods, we use maximum likelihood estimation and 2-D MCMC likelihood analyses with multivariate likelihood models to estimate the parameter biases due to the skewed likelihood distributions of shear correlation functions. In both analyses, we do not observe significant parameter biases in terms of $\Omega_{\rm m}$ and $\sigma_{\rm 8}$ due to the Gaussian likelihood assumption in our data. Note that the results of the Edgeworth likelihood model provide a lower limit on the bias due to non-Gaussian likelihood functions for the 100 deg$^2$ simulations since this model is built from marginal Edgeworth corrections of simple Gaussian functions in the PCA space based on the assumption of independence of PCs and the marginal distributions of PCA components exhibit small skewness and kurtosis. The multivariate non-parametric models that are not sensitive to coordinate rotations provide more general constraints on the bias, but their performance in terms of predicting hold-out data does not exceed that of the Edgeworth model given our data. We are not asserting that this is the final conclusion for the parameter biases due to the uncertainty of likelihoods. Measuring the high-dimensional likelihoods is extremely difficult. It is likely that our likelihood models, especially the non-parametric models, would benefit from orders of magnitude of increase in the number of simulation realizations. However, given the currently available number of simulation realizations, our results show no significant biases using the standard multivariate Gaussian likelihood.

In Appendices \ref{sec:third_moment} and \ref{mean_mode} we show that the skewness and the $(\rm {mode-mean})/\sigma$ of shear correlation functions decrease with the survey area as $f_\text{sky}^{-1/2}$. The non-Gaussianity of the likelihood and the resulting cosmological parameter biases would therefore be even smaller for large surveys such as LSST. 
The scaling relation is derived by assuming that the shear fields are Gaussian, corresponding to larger scales. Future work should explore the relative contributions of the non-linear growth (non-Gaussian field) to the non-Gaussianity in likelihoods.

Therefore, given the small bias measured from the current simulations, we do not expect noticeable biases due to the non-Gaussian distribution of the weak lensing shear correlation function for future generation of large-scale structure observations. A multivariate Gaussian likelihood will continue to be a valid approximation in cosmic shear analyses. 

In addition to our findings on the Gaussian approximation to the likelihood function, PCA poses a straightforward avenue to solving some of the practical problems related to covariance matrices. Several practical challenges for debiasing the covariance matrix of the weak lensing observables from simulations have been featured in the literature \citep{ 2011ApJ...737...11S, 2013PhRvD..88f3537D, 2014MNRAS.442.2728T,2016MNRAS.458.4462B},  all of which are connected to the large number of data points expected in future cosmic shear surveys.   

One way to address the problem is to identify a more efficient observable and hence compress the cosmological information into fewer data points. We have shown that the assumption of a multivariate Gaussian in PCA coordinates is valid for the mock data and that PCA ranks the data points efficiently as a function of signal-to-noise. This allows us to reduce the dimensionality of the data vector by throwing out modes with low signal-to-noise, which in turn alleviates some of the pressing problems related to covariance inversion. 

Data compression approaches have been suggested in the past, such as MOPED \citep{doi:10.1046/j.1365-8711.2000.03692.x, stx2326} and COSEBIs \citep{2010A&A...520A.116S}; it remains to be seen if the likelihood function of these summary statistics is closer to a Gaussian than, e.g. for the shear two-point correlation function. Using PCA, we have been able to reduce the number of dimensions from 770 to 40 for data without shape noise and from 770 to 150 for data with shape noise with acceptable loss in cosmological information.

In brief, our conclusions are itemized as follows:

\begin{enumerate}
    \item We find significant non-Gaussianity in the marginal distributions of the correlation functions, indicating that the multivariate likelihood distributions are non-Gaussian.
    \item We estimate the bias in cosmological parameters induced by ignoring the non-Gaussianity of the likelihood with maximum likelihood estimation assuming a Gaussian likelihood, and do not detect strong biases in $\Omega_{\rm m}$ and $\sigma_{\rm 8}$ in our simulated data.
    \item Since the $(\rm {mode-mean})/\sigma$ of shear correlation functions scales with the survey area as $f_\text{sky}^{-1/2}$ under the assumption that the fields are Gaussian and that the angular power spectra follow a gamma distribution, we expect the bias due to assuming a Gaussian likelihood to be smaller for LSST than for these small-area mock catalogs.
    \item Fully reconstructing high-dimensional distributions directly from simulations is very difficult, requiring simulation volumes well beyond those presently available for this study. Our results based on the described set of simulated data suggest that neglecting the non-Gaussianity of the likelihood for shear-shear correlations is not a significant source of bias for ongoing surveys or even future ones such as LSST. 
\end{enumerate}

While the simulated weak lensing data implies that the impact of non-Gaussianity is negligible in current and future cosmic shear surveys, it is an open question if this result is stable for different redshift distributions, different survey parameters, and different cosmologies. In \cite{2009A&A...502..721E} they show that the covariance and hence the likelihood is cosmology-dependent and that the strength of this effect depends on the specific properties of the survey. It is also unclear if some of the systematics that affect shear observables can introduce non-Gaussianity (e.g., as foregrounds do in case of the CMB), which could warrant further mitigation strategies. Finally, it will be interesting to explore the non-Gaussianity of the likelihood function for the multi-probe analysis case, e.g., when including galaxy clustering, galaxy-galaxy lensing, and perhaps even higher-order statistics of auto and cross observables of clustering and shear. A multi-probe simulation with even more realizations is required for the joint data vector.

\section*{Acknowledgments}

This paper has undergone internal review in the LSST Dark Energy Science Collaboration. The authors would like to give thanks to the DESC Publication Board and the internal reviewers A.~Heavens, R.~Rosenfeld and J.~Zuntz for the feedback and comments that improved our manuscript.
CHL, JHD, TE and RM developed the concept and designed the analysis framework; CHL carried out the analysis and prepared the manuscript with the help of  JHD, TE, RM, ABL, TP and SS; JHD conducted weak-lensing simulations; TP and ABL designed and performed the statistical model assessment, and fitted the nonparametric likelihood models; SS derived expressions of the third moment of shear and mean-mode difference.
JHD acknowledges support from the European Commission under a Marie-Sk{\l}odowska-Curie European Fellowship (EU project 656869) and from the European Research Council under grant number 647112.
Computations for the $N$-body simulations were performed in part on the Orcinus supercomputer at the WestGrid HPC consortium (www.westgrid.ca), 
in part  on the GPC supercomputer at the SciNet HPC Consortium. SciNet is funded by: the Canada Foundation for Innovation under the auspices of Compute Canada;
the Government of Ontario; Ontario Research Fund - Research Excellence; and the University of Toronto. 
This work was partially supported by NSF DMS-1520786,  DOE grant DESC0011114, and Department of Energy Cosmic Frontier program. 
Part of the research was carried out at the Jet Propulsion Laboratory, California Institute of Technology, under a contract with the National Aeronautics and Space Administration and is supported by NASA ROSES ATP 16-ATP16-0084 grant and by NASA ROSES 16-ADAP16-0116.

The DESC acknowledges ongoing support from the Institut National de Physique Nucl\'eaire et de Physique des Particules in France; the Science \& Technology Facilities Council in the United Kingdom; and the Department of Energy, the National Science Foundation, and the LSST Corporation in the United States.  DESC uses resources of the IN2P3 Computing Center (CC-IN2P3--Lyon/Villeurbanne - France) funded by the Centre National de la Recherche Scientifique; the National Energy Research Scientific Computing Center, a DOE Office of Science User Facility supported by the Office of Science of the U.S.\ Department of Energy under Contract No.\ DE-AC02-05CH11231; STFC DiRAC HPC Facilities, funded by UK BIS National E-infrastructure capital grants; and the UK particle physics grid, supported by the GridPP Collaboration.  This work was performed in part under DOE Contract DE-AC02-76SF00515.

\section*{Data availability}
The data generated and analyzed in this article are available from the corresponding author upon reasonable request.
\bibliography{ng}
\appendix
\onecolumn
\section{Third  Central Moment}
      	\subsection{Convergence}	
        \label{sec:third_moment}
			Here we derive the expression for the third central moment of the (tomographic) 
            cross-correlation function of
            two convergence maps denoted by $X$ and $Y$. 
            Similar calculations have also been done by \cite{Keitel2011}. We use a different notation that is 
            easier to generalize for 
            cross-correlations and restrict ourselves to computing the 
            third central moment, though calculations can be extended straightforwardly but tediously to 
            higher moments. We show here how skewness is affected by $f_{sky}$, a relation we used in Sect. \ref{sec:skewness} in the main text.
            
            For simplicity, we will work in the flat-sky approximation.
            We will also assume that the fields are Gaussian and ignore the 
            non-Gaussian terms, though we will indicate 
            where the non-Gaussian terms should enter. Using simulations, we 
            have already shown that at the small scales, where the non-Gaussian terms 
            are most important, the impact
            of non-Gaussian likelihood is small.

    		 The cross-correlation function of two fields can be written as 
			 \begin{align}
    		 	{\hxi}_{XY}(\vec \theta)
							=&\frac{1}{\mathcal A_W(\vec \theta)}\int d^2\vec \theta'W(\vec \theta'+\vec \theta)W(\vec 
                            \theta')
							\hdelta_1(\vec \theta')\hdelta_2(\vec \theta'+\vec \theta)
    		 \end{align} 
			where $\kappa_1$ belongs to field $X$ and $\kappa_2$ to field $Y$. 
			$W(\vec \theta)$ is the survey window function. We have assumed that the noise and the $\kappa_i$ have 
			zero mean and are also uncorrelated with each other on all scales. 
			The normalization factor is the integral over window functions
			\begin{align}
				\mathcal A_W (\vec \theta)=\int d^2\vec \theta' W(\vec \theta'+\vec \theta)W(\vec \theta')
				=\int\frac{ d^2\vec \ell }{(2\pi)^2}e^{-\mathrm{i}\vec \ell \cdot\vec \theta}\tilde W(\vec \ell )\tilde W(-\vec \ell )
				=\int\frac{ d\vec \ell\,\ell }{2\pi}J_0(\ell\theta)\tilde W(\ell )\tilde W(- \ell )
			\end{align}
			
			The third central moment ($S_3$) of the correlation function is given as
			
			\begin{align}
	\text{S}_3(\widehat\xi_{\kappa_1\kappa_2}(\vec \theta_i)\widehat\xi_{\kappa_3\kappa_4}(\vec 
                \theta_j)\widehat\xi_{\kappa_5\kappa_6}(\vec \theta_l)) 
                &= \mean{\left( \widehat\xi_{\kappa_1\kappa_2}(\vec \theta_i) - \mean{\widehat\xi_{\kappa_1\kappa_2}(\vec \theta_i)} \right) \left( \widehat\xi_{\kappa_3\kappa_4}(\vec \theta_j) - \mean{\widehat\xi_{\kappa_3\kappa_4}(\vec \theta_j)} \right)\left( \widehat\xi_{\kappa_5\kappa_6}(\vec \theta_l) - \mean{\widehat\xi_{\kappa_5\kappa_6}(\vec \theta_l)} \right) } \nonumber\\
                &=
					\mean{\hxi_{\kappa_1\kappa_2}(\vec \theta_i)\hxi_{\kappa_3\kappa_4}(\vec 
                    \theta_j)\hxi_{\kappa_5\kappa_6}(\vec \theta_l)}
                    +2\mean{\hxi_{\kappa_1\kappa_2}(\vec \theta_i)}\mean{\hxi_{\kappa_3\kappa_4}(\vec 	
                    \theta_j)}\mean{\hxi_{\kappa_5\kappa_6}(\vec \theta_l)}\nonumber\\
					& \quad -\left\{\mean{\hxi_{\kappa_1\kappa_2}(\vec \theta_i)}\text{Cov}(\hxi_{\kappa_3\kappa_4}(\vec \theta_j),\hxi_{\kappa_5\kappa_6}(\vec \theta_l))+\text{perms}\right\}
				\label{eq:multi-term}
			\end{align}
			
			where $\kappa_1,\kappa_3,\kappa_5$ belong to field $X$ and $\kappa_2,\kappa_4,\kappa_6$ belong to $Y$. $\text{S}_3$ is a function of three $\theta$ variables and is therefore a third-order tensor. The 
            `perms' denotes permutations over combinations of $(12,34,56)$, where one combination gives $\xi$ and other two 
            give covariance. So there are 6 permutations in the last term of Eq.~\eqref{eq:multi-term} (or 3 terms if the symmetry of covariance is considered).
			We use $\mean{\kappa_1\kappa_2\kappa_3\kappa_4\kappa_5\kappa_6}_{ijl}$ as short-hand for the first term on 
            the right hand side in
            Eq.~\eqref{eq:multi-term},
			which we would like to simplify:
			\begin{align}
				\mean{\kappa_1\kappa_2\kappa_3\kappa_4\kappa_5\kappa_6}_{ij}=&\frac{1}{\mathcal A_W(\vec \theta_i)\mathcal A_W(\vec \theta_j)\mathcal A_W(\vec \theta_l)}
				\int d^2 \vec \theta \int d^2 \vec \theta' \int d^2 \vec \theta'' 
				\kappa_1(\vec{\theta}) \kappa_2(\vec{\theta}+\vec{\theta}_i) \kappa_3(\vec \theta')\kappa_4(\vec{\theta}'+\vec{\theta}_j)\kappa_5(\vec \theta'')\kappa_6(\vec{\theta}''+\vec{\theta}_l)
				\nonumber\\
				&W(\vec \theta)W(\vec \theta')W(\vec \theta+\vec \theta_i)W(\vec \theta'+\vec \theta_j)W(\vec \theta'')W(\vec \theta''+\vec \theta_l)
			\end{align}
			
			Writing the $\kappa_i$ in terms of its Fourier space counterpart $\tilde \kappa_i$, we get
			\begin{align}
				\mean{\kappa_1\kappa_2\kappa_3\kappa_4\kappa_5\kappa_6}_{ij}=&\frac{1}{\mathcal A_W(\vec \theta_i)\mathcal 
				A_W(\vec \theta_j)\mathcal A_W(\vec \theta_l)}
				\int d^2 \vec \theta \int d^2 \vec \theta'\int d^2 \vec \theta''
				\iiiint \prod_{n=1}^6\left[\frac{d^2 \vec \ell _n}{(2\pi)^2}\right]
				\iiiint\prod_{m=1}^6\left[\frac{d^2 \vec q_m}{(2\pi)^2}\tilde W(\vec q_m)\right]\nonumber\\&
				\times
				e^{i(\vec{\ell}_1-\vec{q}_1)\cdot\vec{\theta}}
				e^{i(\vec{\ell}_2-\vec{q}_2)\cdot(\vec{\theta}+\vec{r_i})}
				e^{i(\vec{\ell}_3-\vec{q}_3)\cdot\vec \theta'}e^{i(\vec{\ell}_4-\vec{q}_4)\cdot(\vec{\theta}'+\vec{\theta}_j)}
				e^{i(\vec{\ell}_5-\vec{q}_5)\cdot\vec \theta''}e^{i(\vec{\ell}_6-\vec{q}_6)\cdot(\vec{\theta}''+\vec{\theta}_l)} \\
                &\times \mean{\tilde \kappa_1(\vec \ell _1)\tilde \kappa_2(\vec \ell _2)
				\tilde \kappa_3(\vec \ell _3)\tilde \kappa_4(\vec \ell _4)\tilde \kappa_5(\vec \ell _5)\tilde \kappa_6(\vec \ell _6)\nonumber}
				\\
				\mean{\kappa_1\kappa_2\kappa_3\kappa_4\kappa_5\kappa_6}_{ij}=&\frac{1}{\mathcal A_W(\vec \theta_i)\mathcal A_W(\vec \theta_j)\mathcal A_W(\vec \theta_l)}
				\iint \frac{d^2 \vec \ell _1}{(2\pi)^2}\frac{d^2 \vec \ell _3}{(2\pi)^2} \frac{d^2 \vec \ell _5}{(2\pi)^2} 
				\iiiint\prod_{m=1}^4\left[\frac{d^2 \vec q_m}{(2\pi)^2}\tilde W(\vec q_m)\right]
				e^{-\mathrm{i}(\vec{\ell}_1-\vec{q}_1)\cdot\vec{\theta}_i}e^{-\mathrm{i}(\vec{\ell}_3-\vec{q}_3)\cdot\vec{\theta}_j}
				e^{-\mathrm{i}(\vec{\ell}_5-\vec{q}_5)\cdot\vec{\theta}_j}
				\nonumber\\&\times
				\mean{\tilde \kappa_1(\vec \ell _1)\tilde 
				\kappa_2(-\vec \ell _1+\vec q_1+\vec q_2)\tilde \kappa_3(\vec \ell _3)\tilde \kappa_4(-\vec \ell _3+\vec q_3+\vec q_4)
				\kappa_5(\vec \ell _5)\tilde \kappa_6(-\vec \ell _5+\vec q_5+\vec q_6)}.
				\label{eq:kappa_fourier_space}
			\end{align}
			We have integrated over $d^2 \vec \theta$, $d^2 \vec \theta'$, $d^2 \vec \theta''$ and then over $d^2 \vec \ell _2$, 
			$d^2 \vec \ell _4$, $d^2 \vec \ell _6$ to 
			obtain the last expression.
			
			We now expand the six-point function into two separable parts: the connected or non-Gaussian component
			$\mean{\tilde \kappa_1\tilde \kappa_2\tilde \kappa_3\tilde \kappa_4\kappa_5\kappa_6}'$ 
			and the Gaussian component, which 
			using Wick's theorem can be expanded as the sum of the product of two-point functions:
			\begin{align}
				\mean{\kappa_1\kappa_2\kappa_3\kappa_4\kappa_5\kappa_6}_{ij}
				=&\frac{1}{\mathcal A_W(\vec \theta_i)\mathcal A_W(\vec \theta_j)\mathcal A_W(\vec \theta_l)}
				\iint \frac{d^2 \vec \ell _1}{(2\pi)^2}\frac{d^2 \vec \ell _3}{(2\pi)^2} \frac{d^2 \vec \ell _5}{(2\pi)^2} 
				\iiiint\prod_{m=1}^6\left[\frac{d^2 \vec q_m}{(2\pi)^2}\tilde W(\vec q_m)\right]
				e^{-\mathrm{i}(\vec{\ell}_1-\vec{q}_1)\cdot\vec{\theta}_i}e^{-\mathrm{i}(\vec{\ell}_3-\vec{q}_3)\cdot\vec{\theta}_j}
				e^{-\mathrm{i}(\vec{\ell}_5-\vec{q}_5)\cdot\vec{\theta}_l}
				\nonumber\\&
				\left[
				\mean{\tilde \kappa_1\tilde \kappa_2\tilde \kappa_3\tilde \kappa_4\kappa_5\tilde \kappa_6}'+
				\mean{\tilde \kappa_1\tilde \kappa_2}\mean{\tilde \kappa_3\tilde \kappa_4}\mean{\tilde \kappa_5\tilde \kappa_6}+
					\text{all perms}
				\right]
				\label{eq:general_cov_long}
			\end{align}
            where the `all perms' now denote all possible combinations over $(1,2,3,4,5,6)$. 
			Note that a similar expansion is also required for four-point functions in the covariance terms in 
            Eq.~\eqref{eq:multi-term} 
            (see for example \citealt{Singh2017cov} for a covariance expansion in similar notation).
			Simplifying, the auto-correlation terms from 
            the same theta bin, e.g.\ $\mean{\kappa_1\kappa_2}$, cancel out
			and we are left with the permutations that contain only cross-correlation terms involving at most one $
			\kappa_i$ from each $\vec \theta_j$ bins (there are $^4C_1\times {}^2C_1=8$ such permutations). Thus we have
			
			\begin{align}
				\text{S}_3=&\frac{1}{\mathcal A_W(\vec \theta_i)\mathcal A_W(\vec \theta_j)\mathcal A_W(\vec \theta_l)}
				\iint \frac{d^2 \vec \ell _1}{(2\pi)^2}\frac{d^2 \vec \ell _3}{(2\pi)^2} \frac{d^2 \vec \ell _5}{(2\pi)^2} 
				\iiiint\prod_{m=1}^6\left[\frac{d^2 \vec q_m}{(2\pi)^2}\tilde W(\vec q_m)\right]
				e^{-\mathrm{i}(\vec{\ell}_1-\vec{q}_1)\cdot\vec{\theta}_i}e^{-\mathrm{i}(\vec{\ell}_3-\vec{q}_3)\cdot\vec{\theta}_j}
				e^{-\mathrm{i}(\vec{\ell}_5-\vec{q}_5)\cdot\vec{\theta}_l}
				\nonumber\\&
				\left[
				\mean{\tilde \kappa_1\tilde \kappa_2\tilde \kappa_3\tilde \kappa_4\tilde \kappa_5\tilde \kappa_6}'+
				\mean{\tilde \kappa_1\tilde \kappa_3}\mean{\tilde \kappa_2\tilde \kappa_5}\mean{\tilde \kappa_4\tilde \kappa_6}+
					\text{all cross perms}
				\right]
				\label{eq:connected}
			\end{align}
			To further simplify the expressions, we will assume that the scales of 
            interest are smaller than the survey 
			window size and ignore the coupling between the window function and the power 
            spectra. Within this assumption, the window function integrals can be 
            carried out without the power spectra to finally give
			\begin{align}
				\text{S}_3=&\frac{\mathcal A_W(\vec \theta_i-\vec \theta_j+\vec \theta_k)}{\mathcal A_W(\vec \theta_i)\mathcal 
				A_W(\vec \theta_j)
				\mathcal A_W(\vec \theta_l)}
				\iint \frac{d^2 \vec \ell }{(2\pi)^2} 
				e^{-\mathrm{i}\vec{\ell}\cdot(\vec{\theta}_i-\vec{\theta}_j+\vec{\theta}_l)}
				\left[
					C_{13}(\ell)C_{25}(\ell)C_{46}(\ell)+
					\text{all cross perms}
				\right]\\
				\text{S}_3=&\frac{\mathcal A_W(\vec \theta_i-\vec \theta_j+\vec \theta_k)}{\mathcal A_W(\vec \theta_i)\mathcal 
				A_W(\vec \theta_j)\mathcal A_W(\vec \theta_l)}
				\int \frac{d \vec \ell \,\ell}{2\pi}J_{0}(\ell\theta_i)J_{0}(\ell\theta_j)J_{0}(\ell\theta_l) 
				\left[
					2C_{XY}(\ell)^3+6C_{XX}(\ell)C_{YY}(\ell)C_{XY}(\ell)
				\right]\label{eq:kappa_moment}
			\end{align}
            where in the second equation we have used the fact that odd-odd 
            combinations such as `13' give the auto-correlation 
            of field $X$, $C_{XX}$, even-even combinations give the auto-correlation of $Y$, $C_{YY}$ and 
            the even-odd combinations give the
            cross-correlation $C_{XY}$.
    Since $A_W$ scales with the fraction of sky covered by a survey, $f_\text{sky}$, 
    Eq.~\eqref{eq:kappa_moment} suggests 
    that the third central moment scales as $f_\text{sky}^{-2}$. 
    Thus for any given scale that 
    is well within the survey window size, the third central moment of the convergence 
    correlation
    function decreases faster than the covariance as the survey area increases.
    However, as the scale $\theta$ approaches the size of survey, the area factor in 
    the normalization approaches zero, i.e.\ $A_W(\theta)\rightarrow0$ for large 
    $\theta$ and $S_3$ will 
    increase. $S_3$ rises faster than covariance, $S_2$ and thus the skewness increases
    at large scales. This is consistent with the expectation from the central limit 
    theorem as large scales have fewer modes within the survey and we expect them to be
    more skewed.
    
    We also note that we have ignored the connected terms, which can be important if the fields are non-Gaussian. The connected terms have three factors of area, $A_W$ in the denominator and a factor of window functions in the numerators.
    Under the assumption that the window and the connected terms can be decoupled, the window term in the numerator will cancel one factor of $A_W$ and the connected term will also scale similarly to the Gaussian term. The decoupling of the window is not well justified in the case of coupling between large-scale (super-sample) and small-scale modes, but such coupling terms have also been shown to scale with similar factors of area in the context of super-sample covariance. We have not tested such scaling for the six-point function in this work.

	\subsection{Shear}
    	For shear we begin by noting that $\xi_+\propto\langle\gamma_{_X}\gamma_{_Y}^*\rangle$ and $\xi_-\propto\langle\gamma_{_X}\gamma_{_Y}\rangle$. Using these relations and the expressions from the previous subsection, we get
        \begin{align}
        \text{S}_3(\xi_-)=&\frac{\mathcal A_W(\vec \theta_i-\vec \theta_j+\vec \theta_k)}{\mathcal A_W(\vec \theta_i)\mathcal 
				A_W(\vec \theta_j)\mathcal A_W(\vec \theta_l)}
				\int \frac{d \vec \ell \, \ell}{2\pi}J_{4}(\ell\theta_i)J_{4}(\ell\theta_j)J_{4}(\ell\theta_k) 
				\left[
					2C_{XY}(\ell)^3+6C_{XX}(\ell)C_{YY}(\ell)C_{XY}(\ell)
				\right]\\
                \text{S}_3(\xi_+)=&\frac{\mathcal A_W(\vec \theta_i-\vec \theta_j+\vec \theta_k)}{\mathcal A_W(\vec \theta_i)\mathcal 
				A_W(\vec \theta_j)\mathcal A_W(\vec \theta_l)}
				\int \frac{d \vec \ell \,\ell}{2\pi}
				\left[
					2C_{XY}(\ell)^3J_{0}(\ell\theta_i)J_{0}(\ell\theta_j)J_{0}(\ell\theta_k)\right]\nonumber\\
                    &+\left[2C_{XX}(\ell)C_{YY}(\ell)C_{XY}(\ell)J_{4}(\ell\theta_i)J_{4}(\ell\theta_j)J_{0}(\ell\theta_k)+\text{cyc}(i-k)+ \text{cyc}(j-k)
				\right]
			\end{align}
where `cyc' denotes the interchanging of order of the Bessel 
functions for different $\theta$. 
We conclude from the equation above that the same scaling with $f_{sky}$ applies to shear as well.

\section{Difference between mean and mode}
\label{mean_mode}
Assuming a unimodal distribution for a random variable $X$, the difference between the most likely value, mode $\tilde X$, and the mean $\overline X$ in terms of the standard deviation $\sigma$ has an upper bound \citep{Johnson1951}
\begin{equation}
	\frac{|\tilde X - \overline X|}{\sigma}\leq\sqrt{3}
\end{equation}
In our main analysis using $100$~deg$^2$ simulations, we have already shown that the difference 
between the mean and mode is much smaller than this bound.

In order to address the question of how this difference scales 
with the survey area, we begin by noting that for angular 
power spectra $C_\ell$, which follow a gamma distribution with 
$\nu\approx(2\ell+1)f_\text{sky}$ degrees of the freedom 
\citep[see e.g.\ ][ for a detailed discussion of the $C_\ell$ likelihood for CMB]{Percival2006}, 
the difference between the mean and the mode is
\begin{equation}
	\tilde C_{\ell}-\overline C_{\ell}\approx-\frac{2}{\nu}\overline C_{\ell}
\end{equation}
With the variance $\sigma_{C_\ell}^2\approx 2(C_{\ell}^2/\nu)$, the difference in terms of $\sigma$ is 
\begin{equation}
	\frac{|\tilde C_{\ell}-\overline C_{\ell}|^2}{\sigma_{C_\ell}^2}\approx\frac{2}{\nu}.
\end{equation}
Here we assume the cosmic variance-dominated regime, but the scaling 
in the shot noise-dominated regime is similar since the shot noise also 
scales as $1/f_\text{sky}$ for a fixed number density of 
tracers.

Since the correlation functions, $\xi_\pm$, are the Hankel transform of $C_{\ell}$ (in the flat sky approximation), the difference between the mean and the mode for the correlation functions is
\begin{equation}
	\tilde \xi -\overline \xi=\int \frac{d\ell\,\ell}{2\pi}J_n(\ell\theta)(\tilde C_{\ell}-\overline C_{\ell})\approx  
    -\frac{2}{f_\text{sky}}\int \frac{d\ell\,\ell}{2\pi}J_n(\ell\theta)\frac{\overline C_{\ell}}{2\ell+1}
\end{equation}
where $n=0$ for $\xi_+$ and $n=4$ for $\xi_-$.

Since the covariance, $\mathbf{C}$, 
of $\xi$ scales as $1/f_\text{sky}$, the difference between the mean and the mode in terms of signal-to-noise ratio scales as
\begin{equation}\label{eq:xi_bias}
	(\tilde \xi -\overline \xi)^T C^{-1}(\tilde \xi -\overline \xi)\propto \frac{1}{f_\text{sky}}
\end{equation}
Eq.~\eqref{eq:xi_bias} also gives the scaling of the bias in the log likelihood (Gaussian) and the parameter covariance with 
$f_\text{sky}$.

\end{document}